\documentclass[12pt]{article}
\usepackage{graphicx}
\usepackage{dcolumn}
\usepackage{tensor}
\usepackage{bm}
\usepackage{float}
\usepackage{color}
\usepackage{listings}
\usepackage{slashed}
\usepackage{amsmath}
\usepackage[usenames,dvipsnames]{xcolor}
\usepackage{amssymb}
\usepackage{amsfonts}
\usepackage{enumitem}
\usepackage{tensor}
\usepackage{mathtools}
\usepackage{slashed}
\usepackage{verbatim}
\usepackage{feynmf}
\usepackage{hyperref}
\usepackage{mathrsfs,hyperref,color,float}
\hypersetup{colorlinks=true, linkcolor=blue}
\usepackage{cancel}
\numberwithin{equation}{section}
\usepackage{ulem}

\newcommand{\eps}{\epsilon}
\newcommand{\mO}{\mathcal{O}}
\newcommand{\be}{\begin{equation}}
\newcommand{\ee}{\end{equation}}
\newcommand{\pader}[2]{\frac{\partial #1}{\partial #2}}

\newcommand{\zbar}{\bar{z}}
\newcommand{\lbar}{\bar{\lambda}}
\newcommand{\mbar}{\bar{\mu}}

\newcommand{\dmpader}[3]{\frac{\partial^2 #1}{\partial #2 \partial #3}}
\newcommand{\dpader}[2]{\frac{\partial^2 #1}{\partial #2 ^2}}

\newcommand{\Dc}{\mathcal{D}_c}
\newcommand{\Deot}{\Delta_{12}}
\newcommand{\Detf}{\Delta_{34}}

\newcommand{\hba}{\bar{h}}
\newcommand{\pbar}{\bar{\partial}}
\newcommand{\ebar}{\bar{\eta}}

\newcommand{\p}{\partial}

\newcommand{\jbar}{\bar{J}}

\newcommand{\phil}{\varphi^{(l)}}
\newcommand{\phill}{\varphi^{(l-1)}}
\newcommand{\dbar}{\mathchar'26\mkern-12mu d}

\begin{document}
\title{\textbf{The Goldstone Equivalence Theorem and AdS/CFT}\vspace{-.2in}} \date{}
\maketitle

\begin{center}
\author{\large \textbf{Nikhil Anand} and \textbf{Sean Cantrell}}
\end{center}

\begin{center}
{\it Department of Physics \& Astronomy, Johns Hopkins University, Baltimore, MD 21218}\\
\noindent\rule{8cm}{0.4pt}
\end{center}

\vspace{.1in}
\abstract{The Goldstone equivalence theorem allows one to relate scattering amplitudes of massive gauge fields to those of scalar fields in the limit of large scattering energies. We generalize this theorem under the framework of the AdS/CFT correspondence.  First, we obtain an expression of the equivalence theorem in terms of correlation functions of creation and annihilation operators by using an AdS wave function approach to the AdS/CFT dictionary.  It is shown that the divergence of the non-conserved conformal current dual to the bulk gauge field is approximately primary when computing correlators for theories in which the masses of all the exchanged particles are sufficiently large.  The results are then generalized to higher spin fields.

We then go on to generalize the theorem using conformal blocks in two and four-dimensional CFTs. We show that when the scaling dimensions of the exchanged operators are large compared to both their spins and the dimension of the current, the conformal blocks satisfy an equivalence theorem.}

\clearpage
\section{Introduction}  
\begin{figure}[b]
\centering
\includegraphics[width=0.8\textwidth]{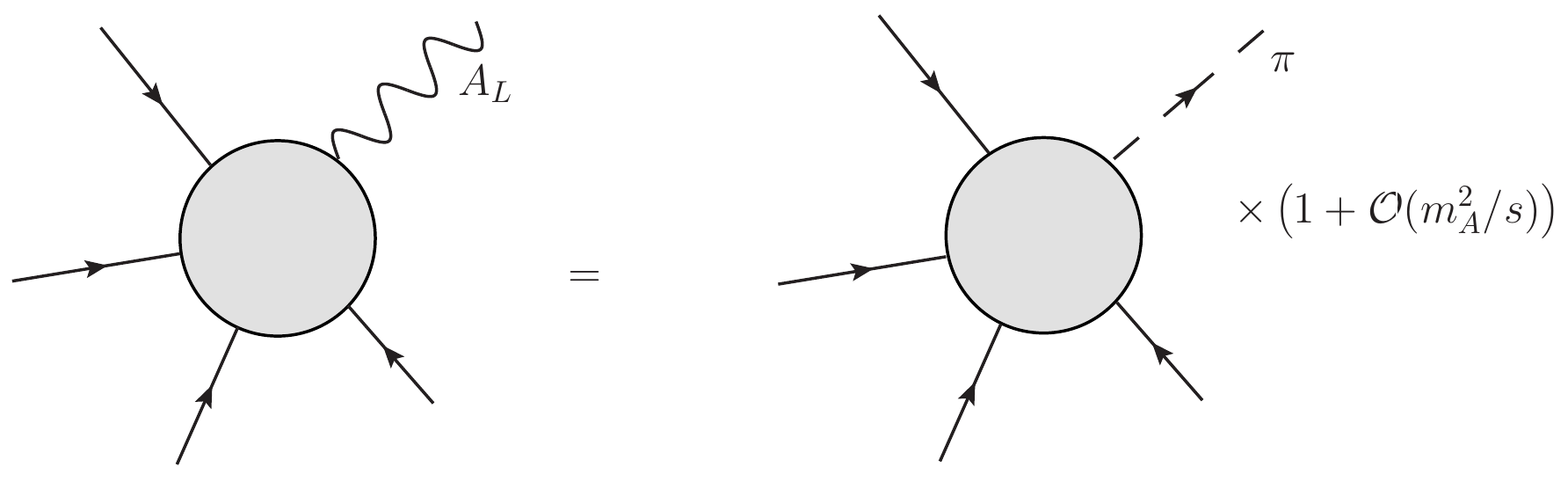}
\caption{\small Diagrammatic representation of the Goldstone equivalence theorem. Here we represent the gauge boson as $W$ in its longitudinally polarized state and its associated Goldstone mode $\phi$. Figure has been adapted from \cite{peskin}.} \label{fig:GET} \end{figure}

The Goldstone equivalence theorem (hereafter `ET') relates the $S$-matrices of processes involving longitudinally polarized massive gauge bosons to the those of scalars when the scattering energies are large compared to the gauge boson's mass \cite{Cornwall}. A diagrammatic depiction of the ET is shown in Fig. \ref{fig:GET}, and the algebraic form is given formally by \begin{align} \label{intro_ET}
S[A_L] \underset{m_A^2/s \to 0}{=} (-i)^n S[\pi]
\end{align} for $n$ replacements of gauge bosons to scalars, $A_L \to \pi$. Here, $m_A$ denotes the boson's mass, $\sqrt{s}$ is the center of mass energy, and $S$ is the $S$-matrix element for some process. The high-energy limit $m_A^2/s \to 0$ corresponds to the massless limit of a spontaneously broken gauge theory, when the Goldstone mode is decoupled.  



A somewhat more sophisticated way of phrasing the ET is to note that the Ward identity for an amplitude involving massive spin-1 gauge fields can be decomposed into a separate gauge field and a scalar mode | these correspond to the 1PI vertex function and the scalar mode created by a current in a spontaneously broken gauge theory, respectively\footnote{This assumes the current is only between on-shell particles. A more careful statement of the ET can be derived for more complicated processes (see, e.g. \cite{chanowitz})}. The ET then implies that the effect of contracting the vertex function with the longitudinal polarization vector reproduces this Ward identity at high energies. This line of analysis can be examined for theories of higher spin massive fields (see, for example, \cite{massivegravity}) and the ET is a strong statement about the limiting behavior of scattering amplitudes in these theories. One need not delve far into the literature to find other uses of the ET. For example, one may go to the unitary gauge to absorb the perturbations of the inflationary scalar field into the graviton component $g_{00}$ \cite{eftinflation}. By the ET, it is possible to write down the effective action that corresponds to the scalar piece that is an excellent approximation at energies much greater than $\sqrt{-\dot{H}}$, taking $H$ to be the canonical Hubble scale.

This suggests that the most immediate consequence of Goldstone equivalency is that in the appropriate regime of validity, one may calculate scalar correlation functions instead of correlation functions which involve gauge fields. Such an application can be very useful, given that gauge field correlation functions typically contain index structures that can render calculations prohibitively difficult. Of course, more indirect and subtle consequences fall out of the ET. For example, the ET implies a nontrivial cancellation of tree level Feynman diagrams involving massive gauge fields of the Standard Model so as to preserve unitarity \cite{peskin}, although this will not be our central focus.

In flat space, the Ward identity for massive gauge bosons and, consequently, the ET are rather straightforward to show.  The Schwinger-Dyson equations provide something akin to a position space statement of the Ward identity by relating the position space vertex functions (amputated Green functions) for processes involving longitudinally polarized gauge bosons to their associated Goldstone bosons ($\Gamma^{\mu}_A$ and $\Gamma_{\pi}$, respectively): 
\begin{align}
\partial_\mu \Gamma^\mu_A (x) \propto \Gamma_\pi (x).
\end{align}

The LSZ formula then provides a map from position space correlation functions to $S$-matrix elements.  The longitudinal polarization vector for large scattering energies is known to be approximately the momentum of the particle: $\epsilon_L ( p ) \approx m_A^{-1} p_\mu + \mO\left( \frac{m_A^2}{E^2} \right)$.  Therefore, one can replace the polarization vector of the incoming vector particle in an $S$-matrix calculation with its momentum.  Pulling the momentum into the LSZ integral and noting that the wave functions appearing in the formula are of the form $e^{i p \cdot x}$, the momentum can be recast in position space as a derivative and the typical Ward identity is found via the Schwinger-Dyson relations:
\begin{align}
p_\mu \Gamma_A^\mu (p) \sim \int d^{d+1} x \, e^{i p \cdot x}  \partial_\mu \Gamma^\mu_A (x) \sim  \int d^{d+1} x \, e^{i p \cdot x}   \Gamma_\pi (x) = \Gamma_\pi (p).
\end{align}
Thus, vector legs with large momentum can be replaced by their associated Goldstone boson. In the center of mass frame, large $s$ implies all external legs have large momentum, and all vector legs can then be replaced with scalar legs.  Since the result is a Lorentz invariant statement, the ET is recovered in all frames.

Since the ET applies at large scattering energies, we trivially expect a similar proof to hold in AdS as one may obtain the flat space limit by sending the bulk curvature scale to zero, which precisely corresponds to bulk interactions where center of mass energies are ``large". However, our goal is less trivial as we will derive a sharp statement of the ET through the direct analysis of (1) massive AdS gauge fields and (2) dual CFT currents. 

The first is achieved by generalizing the usual AdS/CFT dictionary (\cite{Harlow}, \cite{wittenads}, \cite{MaldacenaAdS}, \cite{Gubser:1998bc}).  Our goal is to obtain a similar proof of the ET by extracting descendant vector states and then comparing the result to primary scalar states via the LSZ formula and Schwinger-Dyson relations as above. However, there are several ambiguities that one must address first. For instance, LSZ and Schwinger-Dyson imply an individual vector with large momentum can be replaced by a scalar.  The Lorentzian generalization of this statement is the ET and relies on the fact that, at least in the center of mass frame, large $s$ implies large $\vec{p}$. In a tree level diagram, this means that large exchanged masses imply large $s$.  In AdS, what is the analog to large $\vec{p}$, and what do large exchanged masses in AdS require of the quantum numbers of the external states for the exchange to occur? 

At leading order when a heavy particle is exchanged, the crux of the result can be stated in momentum space, in analogy to Eq. (\ref{intro_ET}), as
\begin{align} 
S[A_z]  =  \left[ - \sqrt{(\Delta_J - ( d - 1) ) (\Delta_J - 1)} \right]^N \left( \prod_{i=1}^N |p_i|^{-1} \right) S[\pi] \label{ads_et2}
\end{align} 
when 
\begin{align}
\frac{\Delta_J}{\Delta} \ll 1, \label{large_exchange_limit}
\end{align}
where $\Delta_J$ is the scaling dimension of the gauge bosons, and $\Delta$ is the scaling dimension of the exchanged particle. The functions $S[O]$ are defined by
\begin{align}
S[O] =& \left( \prod_{i=1}^N \int d^d x_i \int_0^\infty dz_i \, \sqrt{g(z_i)} \, f^{(i)}(p_i; x_i, z_i) \right) \cdot \Gamma_O (\{x_i, z_i\}),
\end{align}
where $f^{(i)}(p_i)$ is the mode function associated with the $i$th external particle and $\Gamma_O$ is the vertex function mediating the interaction.  Generally, $f$ and $\Gamma$ can have index structure.  Under the AdS/CFT dictionary, $S[O]$ may be interpreted as the Fourier transform of the conformal correlators of the operators dual to the bulk fields.

The analogue to the large energy limit for a given gauge boson is (in units where the AdS curvature scale is set to unity)
\begin{align}
\frac{\Delta_J}{s} \ll 1, \label{eqn:helim}
\end{align}
where $s$ can be defined in terms of boundary momenta. 

Massive gauge bosons in AdS are dual to boundary currents that are not conserved.   Conserved currents have vanishing divergence, which implies that certain descendant states are eliminated from the Hilbert space of the theory.  In other words, they belong to the ``short" representation of the conformal algebra, as there are fewer states than those associated with the ``long'' representation of a non-conserved current (\cite{Minwalla:1997ka}, \cite{Beisert:2004ry}). One may combine short representations to obtain a long representation \cite{Bhattacharya:2008zy}, which is analogous to supplying additional degrees of freedom to massless bulk gauge fields to make them massive. For a non-conserved current $J$, the limit in Eq. (\ref{eqn:helim}) reproduces the Ward identity at the boundary,
\begin{align}
\partial \cdot \langle J \dots \rangle \propto \langle \mO_\pi \dots \rangle
\end{align}
for a scalar primary $\mO_\pi$.  We find that the naively descendant operator, $\partial \cdot J$, is approximately primary when computing conformal correlators.  Moreover, the non-conserved current is approximately a functional of a primary scalar,
\begin{align}
J_\mu \propto \partial_\mu \partial^{-2} \mO_\pi.
\end{align}

We generalize this result to show the dominant contribution to massive spin-$l$ interactions comes from lower spin Goldstone-like fields.  At sufficiently high energies, the tower of equivalences collapses to leave a dominant scalar contribution.  On the CFT side, this takes the form
\begin{equation}
\mO_{\mu_1 \dots \mu_l} \approx  \partial_{\mu_1} \dots \partial_{\mu_l} \left( \partial^{-2} \right)^l \mO.
\end{equation}

The above results come from the \textit{bulk} ET. It is a natural continuation, then, to see if the ET can be extracted purely in terms of CFT operators and correlation functions, without any reference to the specifics of the bulk interacting theory. Since conformal field theories do not admit an $S$-matrix in the traditional sense, a natural concern might be how the high energy limit is extracted in terms of dual operators. Moreover, the ET is intimately tied with the polarization vectors associated with massive gauge bosons. Are there analogs of flat-space polarization vectors that can be contracted with conformal correlators of tensor currents? In order to make any progress, we will turn to a completely model-independent, bottom-up approach to CFTs in the form of the operator product expansion (`OPE') \cite{Ferrara1973161}, where one may expand in the distance between two operators \be \mO(x)\mO(y) \sim \sum_{\mO} \lambda_\mO  C(x-y,\partial_y) \mO(y).\ee  The sum will generally run over all primary operators present in the theory (i.e. the scaling dimensions $\Delta$ and spins $\ell$ of these states) and the coefficients $\lambda$ of the above algebra specify the dynamics of the theory. The function $C(x-y,\partial_y)$ is completely fixed by conformal invariance, modulo an overall constant. In a CFT, such an algebra has a finite radius of convergence and becomes particularly powerful in the analysis of correlation functions through the bootstrap program (\cite{Rattazzi:2008pe}, \cite{Rychkov:2009ij}, \cite{Caracciolo:2009bx}, \cite{Rattazzi:2010gj}, \cite{Rattazzi:2010yc}, \cite{Vichi:2011ux}). The OPE also implies that one is always able to reduce higher point correlation functions to those fixed by conformal symmetry (\cite{Sotkov1977375}, \cite{Osborn:1993cr}). For example, applying the OPE twice to a four point function of scalars yields  \begin{align} \langle \phi(x_1) \phi(x_2) \phi(x_3) \phi(x_4) \rangle &= \left(\frac{x_{24}^2}{x_{14}^2} \right)^{\frac{\Delta_1-\Delta_2}{2}}\left(\frac{x_{14}^2}{x_{13}^2} \right)^{\frac{\Delta_3-\Delta_4}{2}} \frac{1}{\left(x_{12}^2 \right)^{\frac{\Delta_1 + \Delta_2}{2}}\left(x_{34}^2 \right)^{\frac{\Delta_3 + \Delta_4}{2}}} \sum_{\mO} \lambda_{\mO}^{12}\lambda_{\mO}^{34} G_\mO(u,v) \nonumber \\
&\equiv \sum_{\mO} \lambda^{12}_\mO \lambda^{34}_\mO W_\mO(u,v), 
\end{align} where $\Delta_i$ are the scaling dimensions of the external operators, and $G_\mO$ are the global conformal blocks and denote the contribution of a given exchanged primary and its descendants to the four-point function. In other words, they are the projection of $\mO_{\Delta,\ell}$ onto the four point function. The general form of these conformal blocks was only recently determined by Dolan \& Osborn (\cite{Dolan:2011dv}, \cite{Dolan:2003hv}) in two and four dimensions for external scalar operators. In order to account for the tensor structures inherent in our analysis (since our external operators will be currents), we will turn to the formalism recently developed in \cite{spin} and \cite{blocks}. In this approach, the role of the polarization vectors is most aptly played by auxiliary vectors $Z^{A_i}$ in embedding space that generate conformally invariant scalar correlation functions out of those that manifestly involve operators with spin. The usage of this formalism to compute the conformal blocks is contingent on the assumption that the exchanged operators in the OPE are symmetric and traceless.

\begin{figure}[h]
\centering
\includegraphics[scale=1.0]{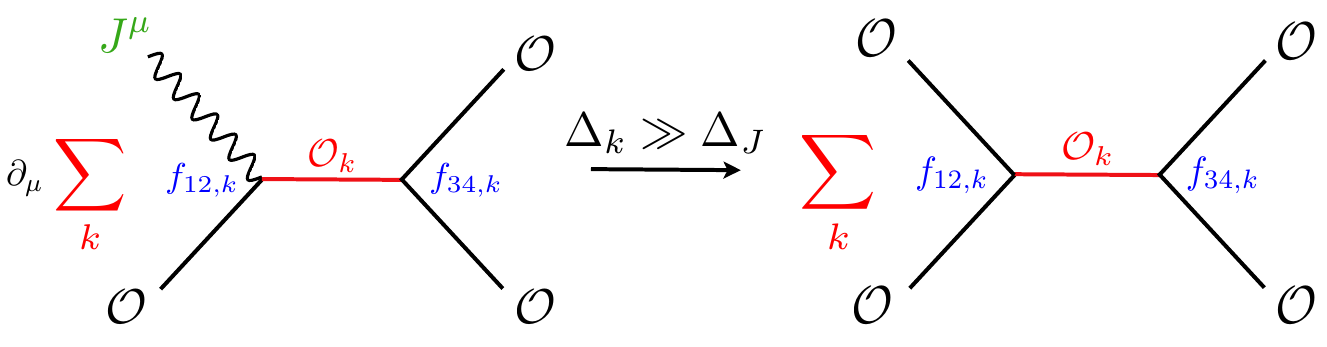}
\caption{\small The CFT ET is schematically illustrated above. The four point function of a spin-1 current and three scalars can be decomposed as a sum over exchanged operators (the conformal blocks). We find that when we the twists of these operators are large compared to the dimension of the current, the blocks associated with the current four point function satisfy an ET (they become scalar blocks, as shown on the right hand side).}
\label{fig:CFTresult}
\end{figure}

Under this caveat, our main result on the CFT side will come from looking at the divergence of a four point function consisting of a single spin-1 non-conserved current and three other scalar operators. We will show that when this correlation function is decomposed in terms of its conformal blocks, the blocks themselves satisfy an ET when the twists $\tau = \Delta - l$ of the exchanged operator are large compared to the dimension of the current (and $\Delta \gg l$). This result is illustrated in Fig. \ref{fig:CFTresult}. We find that this relation holds up to $\mO(1)$ functions that depend only on the coordinates and dimensions of the external operators. As stated, in order to do a partial wave analysis of current correlation functions, we will extensively use two tools: the embedding or ``null cone" formalism and the index-free formalism developed in \cite{spin} and \cite{blocks}, which we will briefly review in \S \ref{sec:spinreview}.

Our paper is organized as follows: in \S \ref{sec:sdrelations}, we review how the Schwinger-Dyson relations work for a general spontaneously broken theory. In \S \ref{sec:etflatspace}, we review how these relations, in conjunction with the LSZ formula, give rise to the familiar ET. In \S \ref{sec:etadsspace}, we then generalize this proof to AdS by expanding gauge fields in terms of mode functions. In \S \ref{higher_spin}, we generalize the AdS ET for arbitrary spin fields. In \S \ref{sec:cftet}, we demonstrate how the ET works in two and four dimensional CFTs; the former is the maximally simple case and serves as a warm-up to the latter, which requires the aforementioned index-free formalism. Finally, we conclude with \S \ref{sec:disc}.

\section{Schwinger-Dyson Relations\label{sec:sdrelations}} \label{lagrangian_sec}
In what follows, it will be necessary to establish the most general form of the Lagrangian for a broken gauge theory in a general spacetime. We will derive a relationship between correlation functions of gauge fields and associated Goldstone fields arising from this Lagrangian. To proceed, we recognize the Goldstone bosons must be derivatively coupled in the Lagrangian; additionally, in the absence of a gauge fixing term, gauge invariance must still be preserved at the Lagrangian level.  This forces the Goldstone sector of the Lagrangian to have the form \begin{align} \mathcal{L}  &\supset  \mathcal{L}_{GS}\left[ (A^a - m_A^{-1} D \pi^a)^2 \right]; \\
\mathcal{L}_{GS} &=  \frac{1}{2} m_A^2 (A^a - m_A^{-1} D \pi^a)^2 + \mathcal{L}_{int} \left[ \frac{1}{2} (A^a - m_A^{-1} D \pi^a)^2 , \chi \right]  \label{lagrangian}
\end{align} where the gauge algebra index, $a$, runs over only the values associated with the broken gauge bosons, $m_A$ is the gauge boson's mass, and $D$ is the gauge covariant derivative.  Here, $\chi$ is used to represent any additional fields that may interact with the gauge and Goldstone bosons.  The interaction Lagrangian, $\mathcal{L}_{int}$, must vanish when fields other than the gauge and Goldstone bosons vanish: $\mathcal{L}_{int}\left[  \frac{1}{2} (A^a - m_A^{-1} D \pi^a)^2,  0 \right] =0$.  To $\mathcal{L}$, we add the $R_\xi$ gauge fixing and ghost terms for full generality \begin{align}
\mathcal{L}_{GF} + \mathcal{L}_{GH} &= - \frac{1}{2} \xi^{-1} G^2 - \bar{c} \frac{\delta G}{\delta \theta} c; \\
G \equiv  g^{M N} \nabla_M A^a_N - \xi m_A \pi^a &\implies 
\frac{\delta G}{\delta \theta} = \nabla^M D_M + \xi m_A,
\end{align} and a generic interaction Lagrangian, $\mathcal{L}_{G, int}( A^a) $, for any additional interactions the gauge fields may have.  This yields the sub-Lagrangian for the broken gauge sector,  \begin{align}
\mathcal{L}&\supset  \left[  -\frac{1}{4} (F^a_{broken})^2 +  \frac{1}{2} m_A^2 (A^a)^2 - \frac{1}{2} \xi^{-1} \left( g^{M N} \nabla_M A^a_N \right)^2 \right] +  \left[ \frac{1}{2} (D \pi^a)^2 - \frac{1}{2} \xi m_A^2 (\pi^a)^2  \right] \nonumber \\
&+  \mathcal{L}_{int} \left[ \frac{1}{2}(A^a - m_A^{-1} D \pi^a)^2 ,  \chi \right] + \mathcal{L}_{G, int}[ A^a] - \bar{c} \frac{\delta G}{\delta \theta} c  . \label{lagrangian2}
\end{align}

As a formality, the Schwinger-Dyson equations for the gauge and Goldstone bosons are given by \begin{align}
[ \nabla_N \nabla^M - \xi^{-1} \nabla^M \nabla_N - ({\nabla^2}^M_N + m_A^2 \delta^M_N) ] \langle T A^{a N} \dots \rangle  &=  \langle T (A^{a M} - m_5^{-1} \partial^M \pi^a)\mathcal{L}'_{int} \dots \rangle \nonumber \\
&+ \langle J^M \rangle + C_G, \label{gauge_eom1}\\
(\nabla^2 + \xi m_A^2) \langle T  \pi^a \dots \rangle =  - m_A^{-1} \nabla_M &\langle T(A^{a M} - m_5^{-1} \partial^M \pi^a) \mathcal{L}_{int}' \dots \rangle \nonumber \\
&+ C_{GS},
\end{align} where the $C$'s are contact terms, $\mathcal{L}'_{int}$ is the derivative of $\mathcal{L}_{int}$ with respect to its first argument, and $J^M$ is a conserved current to which the gauge fields couple, $J^{a M} =  \frac{\p}{\p {A^a}_M} \left[ \mathcal{L}_{G,int}[A^{a}] + \mathcal{L}_{GH}\right]$.  The `$\dots$' include other field operators.

The relevant consequence of these equations for the ET is found by taking the covariant divergence of Eq. (\ref{gauge_eom1}), yielding, up to contact terms,\begin{equation} \label{DS_upshot}
m^{-1}_A \nabla_M[ \nabla_N \nabla^M - \xi^{-1} \nabla^M \nabla_N - ({\nabla^2}^M_N + m_A^2 \delta^M_N) ] \langle T A^{a N} \dots \rangle =  - [\nabla^2  + \xi m_A^2] \langle T  \pi^a \dots \rangle, 
\end{equation} where $T$ is the time-ordering operator.  It is worthwhile noting that $m_A$ here is actually the physical mass, not simply the bare mass that naively appears in the Lagrangian. The significance of the above equation is that there exists projection operators that relate correlation functions involving gauge fields to those involving Goldstone modes. Generally, wave function renormalization factors must also appear, but it has been shown that a renormalization scheme can always be chosen so that they will cancel in Eq. (\ref{DS_upshot}) \cite{bagger_1990}. 

In this section, we made the gauge index explicit.  However, it will not affect any future results, so we will drop it simply as a notational convenience.

\section{ET in Flat Space\label{sec:etflatspace}}
\subsection{Proof for External Legs with Large Momenta\label{flat_et}} 

Consider an arbitrary $S$-matrix element involving a longitudinally polarized gauge boson with spatial momentum $\vec{p}$ and mass $m_A$ in flat space written in terms of correlation functions of fields per the LSZ reduction scheme: 

\begin{equation} \label{scattering1}
\langle \Psi_F | A_L, \vec{p} ; \Psi_I \rangle = i \int d^{d+1} x \, \epsilon_{L, \mu} e^{-i p \cdot x} \left[ \partial^2 \delta^\mu_\nu + m_A^2\delta^\mu_\nu  -(1 - \xi^{-1} )\partial^\mu \partial_\nu \right] \langle T A^{ \nu} \dots \rangle,
\end{equation} 

where $p_0=\sqrt{\vec{p}^2 + m_A^2}$.  The states $\Psi_{I,F}$ are assumed to be created from functions of creation and annihilation operators of various fields, which are included in the `$\dots$' on the RHS of Eq. (\ref{scattering1}).   The differential operator acting on the correlation function can be identified as the one on the LHS of Eq. (\ref{gauge_eom1}).  The contact terms on the RHS contribute to the identity part of the $S$-matrix element.  This identity piece trivially satisfies the ET since it must be the same for scalars and vectors of any masses up to the necessary minus sign for each freely propagating external leg, which arises from the polarization normalization condition $\epsilon_{s}(p) \cdot \epsilon_{s} (-p)=-1$ .  We thus concern ourselves with only the role of interacting processes in the ET and ignore contact terms henceforth.

In the limit $\vec{p}\,^2 \gg m_A^2$, we find $\epsilon_{L, \mu} = m_A^{-1} p_\mu + \mO (m_A^2/\vec{p}^2)$.  
Inserting this into Eq. (\ref{scattering1}) we obtain \begin{align}
\langle \Psi_F | A_L, \vec{p} ; \Psi_I \rangle &=  \int d^{d+1} x \,   e^{-i p \cdot x}  m_A^{-1} \partial_\mu   \left[ \partial^2 \delta^\mu_\nu + m_A^2\delta^\mu_\nu  -(1 - \xi^{-1} )\partial^\mu \partial_\nu \right]\langle T A^{ \nu} \dots \rangle \label{no_surface} \\
&=   \int d^{d+1} x \,   e^{- i p \cdot x}   \left[ \partial^2 + \xi m_A^2 \right]  \langle T \pi \dots \rangle = - i \langle \Psi_F | \pi, \vec{p}; \Psi_I \rangle , \label{goldstone_flat}
\end{align}
where $\langle \Psi_F | A_L, \vec{p} ; \Psi_I \rangle$  is the process involving external gauge bosons and $\langle \Psi_F | \pi^a, \vec{p}; \Psi_I \rangle$ is the $S$-matrix element for the same process in which the longitudinally polarized gauge boson has been replaced with its corresponding Goldstone boson.  The second to last line was obtained from the preceding one by using Eq. (\ref{DS_upshot}).  The equivalence between the last two lines requires that  $p^2 \to 0$.  While the masses of the Goldstone bosons and gauge bosons generally differ, their energies are dominated by momentum and are thus both approximately massless.

Note that the statement $\vec{p}^2 \gg m_A^2$ is not Lorentz-invariant while the equivalency of the $S$-matrices themselves must be.  The appropriate Lorentzian generalization of this frame-dependent limit should be $\frac{m_A^2}{s} \ll 1$, where $s$ is the center of mass energy.  Since $s$ involves the energies and momenta of all the gauge bosons in the scattering process, this suggests we may make the replacement $A \to \pi$ for all longitudinally polarized gauge bosons as long as this limit is satisfied.  To confirm that this is indeed the correct Lorentzian generalization, consider a general scattering process in the center of mass frame with a number of incoming particles, with energies $p_{0,i}=\sqrt{\vec{p}_i^2+ m_i^2}$. In this frame, $ \sum_{i} \vec{p}_i = 0$, so $s= \sum_{j, i} p_{0,i}  p_{0,j}$.  Considering the case in which $s \sim m_A^2$, increasing $s$ by a factor $c$ sufficiently large that $\frac{m_a^2}{s} \ll 1$, each $p_0$ must increase by a factor $\sqrt{c}$.  This can only be accomplished by increasing $| \vec{p}_i |$ by a correspondingly large factor such that $\vec{p}_i^2 \gg m_i^2$. The LSZ formula for this process with $n$ gauge bosons will contain $\epsilon_L(\vec{p}_1) \otimes \epsilon_L(\vec{p}_2) \otimes \dots \otimes \epsilon_L(\vec{p}_n) \approx m^{-n} p_1 \otimes \dots \otimes p_n$. As was shown, the momenta become derivatives of the scalar wave function in the integral in this frame.  This set of derivatives is Lorentz covariant. Transforming to a different frame simply boosts each derivative to a derivative in the new frame, thereby confirming this limit is the correct Lorentzian generalization.

\subsection{The Connection of the Exchange Operator to the ET\label{exchange_op_sec}} 

In the previous section, the ET was demonstrated in the scenario that the center of mass energy is large.  Broken gauge theories clearly admit couplings such that scattering amplitudes have appreciable support at the low energies, in which case the ET is only true for $S$-matrix elements in the high energy limit.  If, instead, the theory contains couplings such that interactions occur only at large energies anyway, then the ET should be automatically satisfied at any momentum scale of the external legs since the identity part of the $S$-matrix trivially satisfies the ET and interactions would be irrelevant until momentum scales that satisfy the ET are reached anyway.  If interactions are mediated by only by particles with mass $m_{\textrm{other}}$ such that $\frac{m_A^2}{m_{\text{other}}^2} \to 0$, then the theory satisfies this coupling criteria.  This follows since poles in the vertex function for scattering processes would be pushed high enough to render interactions negligible except at large energies, $s \sim m_{\text{other}}^2$.

As an example, consider a gauge theory coupled to a Higgs sector with a very large mass and Yukawa couplings to an uncharged scalar:
\begin{align}
\mathcal{L} \supset & - \frac{1}{4} F^2 + | D \Phi(h) |^2 - \frac{1}{2} m_H^2 h^2 - \frac{1}{2} y h \phi^2
\end{align} 
with $\frac{m_A^2}{m_H^2} \ll 1$ and where $\Phi(h)$ is in the fundamental representation of the gauge group.

\begin{figure}[h]
\centering
\includegraphics[scale=0.7]{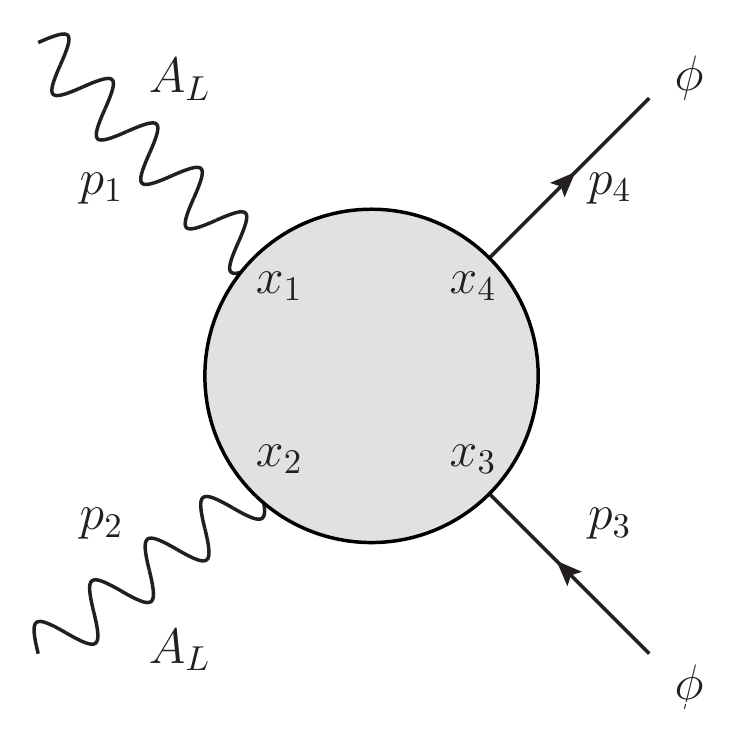}
\caption{\small The depiction of a general process sending two incoming longitudinally polarized gauge bosons to two outgoing scalars.  The vertex function variations $x_1,\dots,x_4$ are integrated over in the LSZ formula.}
\label{fig:flat_blob_exchange}
\end{figure}

This Lagrangian admits the potential four-point interacting process $A_L A_L \to \phi \phi$, which is diagrammatically represented in Fig. \ref{fig:flat_blob_exchange}.
The LSZ formula for this process reads 
\begin{align}
S[A_L] \equiv \langle \vec{p}_3; \vec{p}_4 | \vec{p}_1 , L ; \vec{p}_2, L \rangle &= - \int d^{d+1} x_1 \dots d^{d+1} x_4 \,  \epsilon_{L, \mu_1}(p_1) \epsilon_{L, \mu_2}(p_2) e^{-i (p_1 \cdot x_1 + p_2 \cdot x_2 - p_3 \cdot x_3 - p_4 \cdot x_4)} \nonumber \\
& \times \Gamma^{ \mu_1 \mu_2}(x_1, x_2, x_3, x_4) \label{flat_lsz}
\end{align} 
where 
\begin{align}
 \Gamma^{\mu_1 \mu_2}(x_1, x_2, x_3, x_4) \equiv& {}_A {D^2}^{\mu_1}_{\nu_1} {}_A {D^2}^{ \mu_2}_{ \nu_2} {}_\phi {D^2}_1 {}_\phi {D^2}_2  \langle T A^{ \nu_1}(x_1) A^{ \nu_2} (x_2) \phi(x_3) \phi(x_4) \rangle \label{flat_vertex}
 \end{align}
is the vertex function (amputated Green function) with
\begin{align}
{}_A {D^2}^{\mu}_\nu \equiv& \left[ (\partial^2 + m_A^2) \delta^\mu_\nu - (1 - \xi^{-1}) \partial^\mu \partial_\nu \right], \\
{}_\phi D^2 \equiv& \left[ \partial^2 + m_\phi^2 \right].
\end{align}  

\begin{figure}[h]
\centering
\includegraphics[scale=0.7]{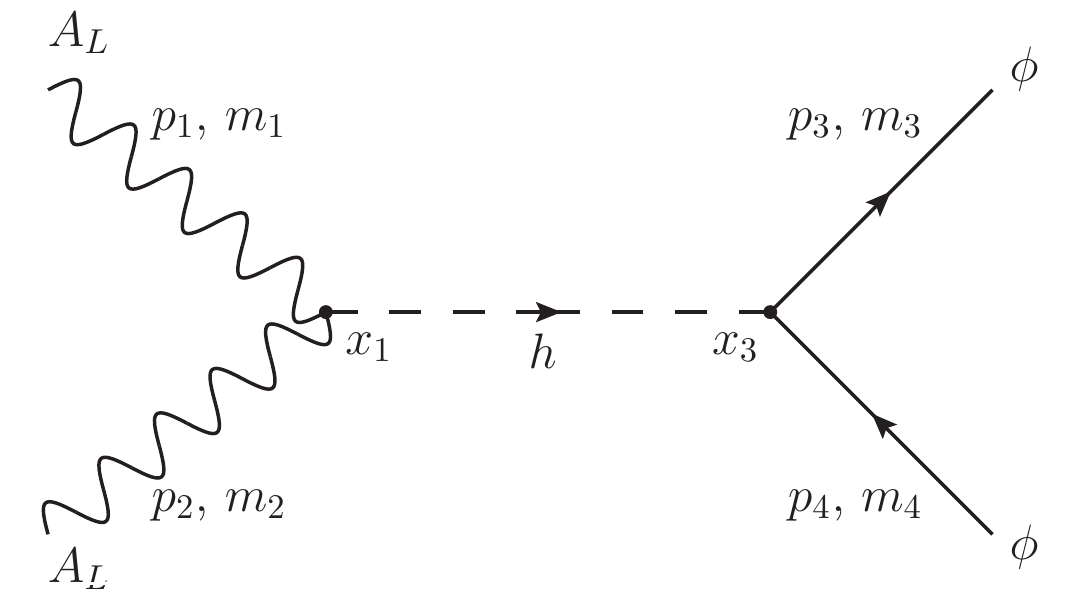}
\caption{\small The leading order contribution to the interacting piece of the $S$-matrix for the process depicted in Fig. \ref{fig:flat_blob_exchange}.}   \label{fig:flat_tree_exchange}
\end{figure}

At the level of interactions, the emergence of the ET for large poles in $\Gamma^{ab, \mu_1 \mu_2}$ is most easily demonstrated at leading order, as shown diagrammatically in Fig. \ref{fig:flat_tree_exchange}. 
with the interaction vertex function given by
\begin{align}
\Gamma^{ \mu_1 \mu_2} \propto   \int d^{d+1} k \frac{i}{k^2 - m_H^2 + i \epsilon} e^{-i k \cdot (x_3 - x_1)}, \label{flat_leading_exchange}
\end{align}

Inserting Eq. (\ref{flat_leading_exchange}) into Eq. (\ref{flat_lsz}) and performing the integrals over $x_1$ and $x_3$ yields a momentum-conserving delta function $(2\pi)^{d+1} \delta^{d+1} (p_1+p_2 - k)$ that forces the constraint $k^2=s$.  For sufficiently large $m_H^2$, the effective operator $A^2 \phi^2$ is extremely suppressed, with scattering amplitude $\mathcal{M} \propto \frac{ y m_A}{m_H^2} \sim 0$, implying that the exchange essentially does not occur, leaving only the identity piece and trivially satisfying the ET.  The exchange becomes relevant when $s \sim m_H^2$, which corresponds to the condition $\frac{m_A^2}{s} \ll 1$, at which point the ET is satisfied anyway.

Evaluating correlation functions of fields involves integrating objects $S[A,\dots]$ over the external momenta, which will generally include some scale for which the ET does not hold.  The above results, however, open the possibility of replacing the longitudinal degree of freedom with the derivative of a scalar for theories that require a large invariant mass to excite exchanges.

\section{ET in AdS\label{sec:etadsspace}}

Expressing $S$-matrix elements using the LSZ formula in previous sections admitted a simple derivation in which the equivalence of the derivative of the scalar wave function and the vector wave function in a particular limit implies the ET from the Schwinger-Dyson equations.  In AdS, the initial and final states which are related by the $S$-matrix are prepared by acting on the vacuum state with CFT operators (\cite{Heemskerk:2009pn}, \cite{Fitzpatrick:2010zm}, \cite{Fitzpatrick:2011jn}). Moreover,  the LSZ formula becomes a generalization of the AdS/CFT dictionary that relates correlators of boundary creation and annihilation operators to integrals of bulk fields. For parallelism and succinctness, these correlators will be referred to as matrix elements since they should reproduce the $S$-matrix in the flat space limit. These matrix elements are related to momentum space conformal correlators, which will be shown within this section and whose utility has been shown in \cite{Raju} and \cite{Penedones}.  By utilizing this generalization, a nearly identical derivation of the ET for matrix elements that follows from the equivalence of wave functions can be made in AdS.

\subsection{AdS Wave Functions}
The LSZ formula for a given external leg in some process in flat space reads schematically as
\begin{align}
S[\phi,i] = \int dx N_i f_i(x) D^2 \langle T \phi \dots \rangle,
\end{align}
where $S$ is the $S$-matrix element involving an external $\phi$ leg labeled by quantum numbers $i$, $f_i(x)$ is the (normalized) wave function for the particle $\phi$ in this process, $N_i$ is the state normalization for that leg, and $D^2$ is a differential operator inverse of the $\phi$ propagator.  The scalar and vector wave functions labeled by momentum are simply $f(\vec{p},x) = \frac{1}{\sqrt{2 p_0}} e^{-i p \cdot x}$ and $h_{s \mu}(\vec{p}, x) = \frac{1}{\sqrt{2 p_0}} \epsilon_{s \mu} (\vec{p}) e^{-i p \cdot x}$, respectively, and the momentum state normalization is given by $| \vec{p}, s \rangle = \sqrt{2 p_0} a^{ \dagger}_s (\vec{p}) | \Omega \rangle \implies N_{\vec{p}} = \sqrt{2 p_0}$.  To obtain an AdS ET using the LSZ formula as a generalization of the AdS/CFT dictionary, we need to specify appropriate AdS scalar and vector wave functions and state normalizations.  In Poincar\'{e} patch coordinates, it is natural to expand the gauge fields as 
\begin{align}
A_M(x,z) = \sum_{s=1}^{d+1} \int d^d \vec{p} \int_0^\infty dm \left[ a_{s}(\vec{p},m) h^\dagger_{s, M} (\vec{p}, m ; x, z) (x) +h.c. \right] \label{vector_expansion} 
\end{align} 
and (real) scalar fields as 
\begin{align} \phi(x,z) = \int d^d \vec{p} \int_0^\infty dm \left[ b (\vec{p},m) f^\dagger (\vec{p}, m; x, z) + h.c. \right] \label{scalar_expansion}
\end{align}
where 
 \begin{align} &f(\vec{p}, m) = \frac{\sqrt{m}}{\sqrt{2 p_{m 0}}} z^{\frac{d}{2}} J_{\Delta_\phi -\frac{d}{2}} (m z) e^{-i p_m \cdot x}, \label{scalar_ads_functions}\\
&h_{s, M} (\vec{p}, m) \underset{s \ne z, \xi}= \left\{ \begin{matrix}
0, & M=z\\
\frac{\sqrt{m}}{\sqrt{2 p_{m 0}}} \epsilon_{s, \mu}(\vec{p}) z^{\frac{d}{2} -1 } J_{\Delta_J - \frac{d}{2}} (m z) e^{- i p_m \cdot x}, & M=\mu
\end{matrix},
\right. \label{vector_ads_functions}\\
&h_{z, M} (\vec{p}, m) = \left\{ \begin{matrix}
\frac{\sqrt{m}}{\sqrt{2 p_{m 0}}} z^{\frac{d}{2}} J_{\Delta_J - \frac{d}{2}} (m z) e^{-i p_m \cdot x}, & M=z \\
- i \frac{\sqrt{m}}{\sqrt{2 p_{m 0}}} \frac{p_{m \mu}}{m^2}     \left[ m z^{\frac{d}{2}} J_{\Delta_J - \frac{d}{2} + 1 } (m z) - [\Delta_J - (d - 1) ] z^{\frac{d}{2} - 1} J_{\Delta_J - \frac{d}{2}} (m z) \right] e^{ - i p_m \cdot x}, & M=\mu
\end{matrix}
\right., \label{z_ads_function}
%
%
\end{align}
in which the particles' masses have been replaced by the their scaling dimensions: $m_{ A}^2 \to [\Delta_J - (d - 1)] (\Delta_J -1)$ and $m_{\phi}^2 \to \Delta_\phi (\Delta_\phi -d)$. The set of vectors $\{\epsilon_{s, \mu}  \}$ are the usual Lorentzian polarizations.  The parameter $p_m$ is just the Lorentzian momentum with mass $|p_m|=m$.  Here, we are letting $m$ be a degree of freedom over which we integrate instead of the traditional $p_0$ typically found in the literature.  This serves two purposes.  First, it naturally imposes the restriction that the boundary momenta be time-like instead of requiring the Fourier transform be only over positive squared norms.  Second, it lends itself to considering a foliation of AdS over space-like Cauchy surfaces at constant Poincar\'{e} patch time, which is more in keeping with the flat space approach.  After our analysis, it should be rather clear that our foliation did not matter and we will ultimately state the results in terms of boundary $d$-momenta, $p$, instead of $\vec{p}$ and $m$.

For completeness, it should be noted that the above listed vector wave functions do not account for the $d+1$ total polarizations that must be summed over in Eq. (\ref{vector_expansion}) to account for all (nominal) degrees of freedom.  There is an additional, unphysical wave function associated with the divergence of $A$ that takes the form $h_{\xi, M} = \partial_M \nabla^{-2} f_{\Delta_\phi \to \Delta_\xi}$ that will be inconsequential in the following sections. It will also be useful to explicitly write the derivative of the scalar wave function: 
 \begin{align}
\partial_M f(\vec{p} , m) =& \left\{ \begin{matrix}
- \frac{\sqrt{m}}{\sqrt{2 p_{m 0}}} \frac{p_{m \mu}}{m^2}     \left[ m z^{\frac{d}{2}} J_{\Delta_J - \frac{d}{2} + 1 } (m z) - \Delta_J z^{\frac{d}{2} - 1} J_{\Delta_J - \frac{d}{2}} (m z) \right] e^{ - i p_m \cdot x}, & M=z \\
- i \frac{\sqrt{m}}{\sqrt{2 p_{m 0}}} p_{m \mu} z^{\frac{d}{2}} J_{\Delta_\phi - \frac{d}{d}} (m z) e^{-i p_m \cdot x}, & M=\mu
\end{matrix}
\right. .
\end{align}

In order to preserve the conformal invariance of the inner products of states, we choose the normalization $| \vec{p},  m, s \rangle = \sqrt{2 p_0  m } \,a^{\dagger}_s (\vec{p} , m) | \Omega \rangle$.  Under this choice in basis wave functions and state normalization, the LSZ-like formula for a matrix element reads
 \begin{align}
S[A,\vec{p},m,s] =& i \int d^4 x \int_0^\infty dz \, N (\vec{p}, m) h_{s, M }(\vec{p}, m)  {}_A {D^{}}^{M}_N \langle T A^{ \nu} \dots \rangle ,\\
S[\phi, \vec{p}, m] =& i \int d^4 x \int_0^\infty dz \,  N (\vec{p}, m) f(\vec{p}, m)   {}_\phi D^2_{x_i}   \langle T \phi \dots \rangle ,
\end{align} 
where 
\begin{align}
{}_A D^{2 M}_N =& \left[  \left({\nabla^2}^M_N + (\Delta_J - (d-1) ) (\Delta_J - 1) \delta^M_N \right) -  \nabla_N \nabla^M +  \xi^{-1} \nabla^M \nabla_N  \right],  \\
{}_\phi D^2 =&  [\partial^2 + \Delta_\phi (\Delta_\phi - d)],
\end{align} 
and $N=\sqrt{2 p_{m 0} m}$.

\subsection{The AdS ET}
Continuing in parallel with the methods from \S \ref{sec:etflatspace}, we must demonstrate the equivalency of the LSZ integrals under the exchange $h_s \to \partial f$ for some particular spin index, `$s$', in some particular limit to show that a given external vector can be replaced with a scalar. In other words, we would like to determine longitudinal polarizations in terms of the mode functions. To understand which spin is relevant, we can briefly consider the dual boundary current and determine which spin corresponds to the degree of freedom introduced by breaking the gauge symmetry in the bulk. To understand what the analogous limit to large momentum in \S \ref{flat_et} is, we can work in the reverse order of the flat space sections and first examine the four-point matrix element in AdS for the same process that was considered in \S \ref{exchange_op_sec} in the limit that the scaling dimension of the exchange operator (i.e. the exchanged mass) is large.  This should then reveal the constraints on the quantum numbers of incoming/outgoing states for the ET to hold.

Since breaking the gauge symmetry generates a mass for the gauge boson, its dual current should not be conserved.  This divergence degree of freedom of the current must thus play the role of the longitudinal degree of freedom in the flat space case, so we should expect that the degree of freedom of the gauge boson that corresponds to the divergence of the current when carried to the boundary is precisely the analogue to the longitudinal degree of freedom.

The gauge boson is identified with its dual current by 
\begin{align}
J^a_\mu &= \lim_{z \to 0} \frac{F^a_{z \mu}}{z^{\Delta_J - 2}} \nonumber \\
&=  \lim_{z \to 0} \frac{\partial_z A^a_\mu - \partial_\mu A^a_z}{z^{\Delta_J - 2}} \nonumber \\
&=  \frac{\Gamma(\Delta_J)}{\Gamma(\Delta_J - 1)}  \lim_{z \to 0} \frac{\bar{A}^a_\mu}{z^{\Delta_J - 1}},
\end{align} 
where 
\begin{align}
\bar{A}^a_M \equiv \left[ \delta^N_M - \partial_M \nabla^{-2} \nabla^N \right] A^a_N
\end{align}
is the gauge field with the gauge dependence projected out.  The replacement $A \to \bar{A}$ can be made since it leaves the field strength tensor unchanged and is useful as it allows us to ignore the $s = \xi$ wave function index in the expansion of $A$.  For notational tractability, we will redefine $J \to \frac{\Gamma(\Delta_J)}{\Gamma(\Delta_J - 1)} J $ to eliminate the gamma factors.  The divergence of the current is then 
 \begin{align}
\partial \cdot J^a &= \lim_{z \to 0} \frac{\partial \cdot \bar{A}^a}{z^{\Delta_J - 1}}\nonumber \\
&= \lim_{z \to 0}  \frac{1}{z^{\Delta_J - 3}} \left[ \nabla_M \bar{A}^{a M} - \nabla_z \bar{A}^{a z} \right] \nonumber \\
&= \lim_{z \to 0}  \frac{1}{z^{\Delta_J - 1}} \left[   z^{d - 1} \partial_z \left( \frac{1}{z^{d - 1}} \bar{A}^{a}_z\right)  \right]. \label{divergence_gauge_relation}
\end{align} 

Since the only wave function that contributes to $\bar{A}_z$ is $h_z$, $s=z$ must play the role of bulk longitudinal polarization.  This result may appear naively gauge dependent since we might expect symmetry breaking to generate a $z$-component only in the $A_z=0$ gauge and not a general $\xi$ gauge. It might also seem unusual that $A_M$ possessing a nontrivial $z$-component seems unrelated to longitudinal propagation in the flat space limit. We know, however, that the longitudinal polarization in flat space is the only polarization with a component in the time direction, so it is not entirely surprising that the relevant wave function is the only one with non-vanishing components in a preferred direction.

\begin{figure}[h]
\centering
\includegraphics[scale=0.7]{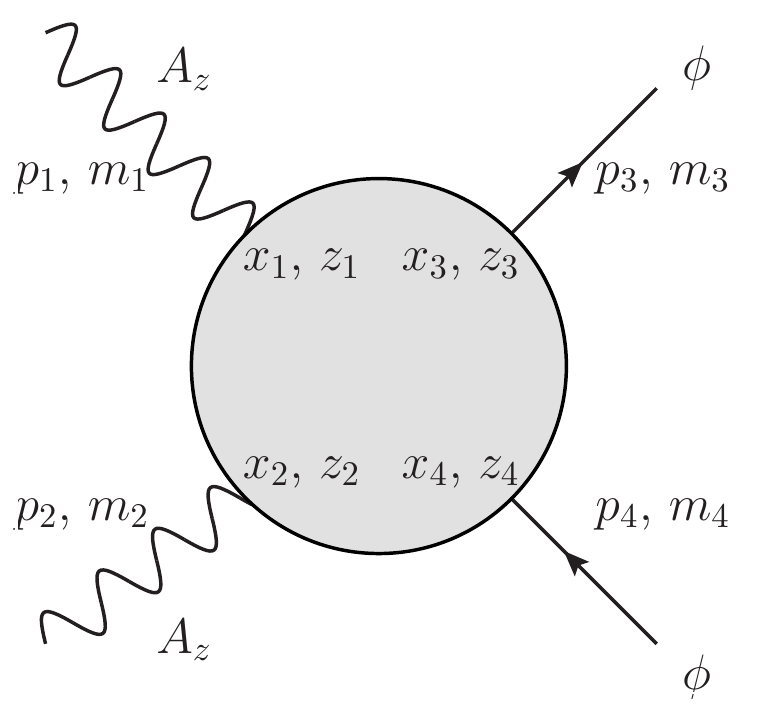}
\caption{\small The general four-point diagram for gauge bosons ``incoming" gauge bosons at different times (as defined by creation operators) ending with ``outgoing" scalars at different times (as defined by annihilation operators). The arguments of the vertex function $\{x_i, z_i\}$ are integrated over in the LSZ formula.}
\label{fig:ads_blob_exchange}
\end{figure}

Having determined the appropriate $s$ index, we now turn to the constraints imposed on the quantum numbers of external states in order to excite a four-point process involving heavy exchanges as depicted in Fig. \ref{fig:ads_blob_exchange}.  The complete LSZ formula for this process is\begin{align}
S[A_z]\equiv &\langle \vec{p}_3,m_3; \vec{p}_4, m_4 | \vec{p}_1, m_1; \vec{p}_2, m_2 \rangle =   \left( \prod_{i=1}^4 \int d^4 x_i \int_0^\infty dz_i  \, \sqrt{ g( z_i)} \right) \nonumber \\
&\times \left( \prod_{i=1}^2 N_A(\vec{p}_i, m_i) h_{z,  M_i} ( \vec{p}_i, m_i; x_i, z_i) \right)    \left( \prod_{i=3}^4 N_\phi(\vec{p}_i, m_i) f_\phi ( \vec{p}_i, n_i; x_i, z_i) \right)   \Gamma^{ M_1 M_2}_{AA\phi\phi} , \label{ads_lsz_exchange}
\end{align} 
where the vertex function, $ \Gamma^{M_1 M_2}_{AA\phi\phi}$, is defined in the analogous way to what was encountered in Eq.(\ref{flat_vertex}).  The important consequence of the Schwinger-Dyson relations now amount to
\begin{align} \label{ds_vertex}
 \nabla_{M_1} \nabla_{M_2} \Gamma^{M_1 M_2}_{AA\phi\phi}  = \left[ (\Delta_J - (d-1) ) (\Delta_J - 1) \right] \Gamma_{ \pi \pi \phi \phi},
\end{align} 
where $\Gamma_{ \pi \pi \phi \phi}$ is the scalar vertex function.  Expectedly, when $\Delta_J =(d-1)$, the gauge boson is massless and the usual Ward identities are satisfied. 
 
As before, the identity part of the matrix trivially satisfies the ET and we turn to the same hypothetical leading order contribution for the exchange to make the notion of large scattering energies concrete.  The vertex function then takes the form 
\begin{align}
\Gamma_{AA\phi\phi}^{ M_1 M_2} \propto  \int_0^\infty dn \, n (z_1 z_3)^{\frac{d}{2}} J_{\Delta -\frac{d}{2}} (n z_1) J_{\Delta - \frac{d}{2}} (n z_3) G_n(x_3 - x_1), \label{leading_ads_exchange}
\end{align} 
where  $G_n$ is the usual lorentzian propagator for a scalar of mass $n$ and $\Delta$ is the scaling dimension of the exchanged scalar.  A heavy/energetic exchange corresponds to large $\Delta$: $\frac{\Delta_J}{\Delta} \ll 1$.  This makes sense since the scaling dimensions of exchanged operators in conformal theories can be thought of as a measurement of the center of mass energy. We will elaborate on this point when we discuss the ET on the side of the CFT.

For $nz \ll \sqrt{\Delta}$, $J_{\Delta -\frac{d}{2}} \approx \frac{1}{\Gamma (\Delta -\frac{d}{2}+1)} \left( \frac{nz}{2} \right)^{\Delta - \frac{d}{2} }$, which is very strongly suppressed by $\Delta$.  So $J_{\Delta - \frac{d}{2}}(n z)$, and consequently the entire LSZ integral\footnote{This follows since any Bessel function $J_\alpha(x)$ dies more quickly as $x \to 0$ than $x \to \infty$}, is dominated by large $n z$ behavior for large $\Delta$.  Then only either $n$ or $z$ needs to be large for the exchange to be relevant.  Since both parameters are integrated over in the LSZ integral, we consider the two relevant regions of parameter space in which one remains finite and the other is large.  

For the first region in which $z$ is finite and $n \to \infty$, the integral in Eq. (\ref{leading_ads_exchange}) is dominated by large $n$. The situation then closely resembles the flat space case: sufficiently large $\Delta$ pushes $n^2$, and consequently the poles in $G_n$, enough to render the exchange negligible except for $s \approx n^2$.  In turn, this requires large $\vec{p}$'s and $m$'s for this part of integration space to contribute to the exchange, which demands that the argument $m z$ in the vector wave functions be large.  If the $|p|$'s are finite, then the contribution of this region of integration space is negligible; if they are large, then $m z \approx \sqrt{ \Delta} \gg \sqrt{\Delta_J}$, and the Bessel functions in the wave functions take their asymptotic forms for large arguments: $J_{\alpha} (m z) \approx \sqrt{\frac{2}{\pi m z}} \cos \left( m z - \frac{\pi}{2} \alpha - \frac{\pi}{4} \right)$.

For the second case in which $n$ is finite and $z \to \infty$, the story is much more trivial. Since $n z \approx \sqrt{\Delta }$ is large, and the poles in $G_n$ set $n = \sqrt{s}$ after integrating over $n$, we conclude $m z \approx \sqrt{ \Delta } \gg \sqrt{\Delta_J }$.  The wave functions then assume the same asymptotic form as the previous case.

To confirm that the scalar and $s=z$ vector wave functions effectively share the same large argument behavior, we explicitly compare the large $m z$ behaviors of $h_{z,M}$ and $\partial_M f$ to find 
\begin{align}
h_{z, M} \underset{m z \gg \sqrt{ \Delta_J - \frac{d}{2} + 1}}{=}& \left\{ \begin{matrix}
\frac{\sqrt{m}}{\sqrt{2 p_{m 0}}} z^{\frac{d}{2}} \sqrt{\frac{2}{\pi m z}} \cos\left[ \frac{\pi}{4} + \frac{\pi}{2} (\Delta_J - \frac{d}{2}) - m z \right] e^{-i p_m \cdot x}, & M=z \\
-i \frac{\sqrt{m}}{\sqrt{2 p_{m 0}}} \left( \frac{p_{m \mu}}{m^2} \right) m z^{\frac{d}{2}} \sqrt{\frac{2}{\pi m z}} \cos\left[ \frac{\pi}{4} + \frac{\pi}{2} (\Delta_J - (\frac{d}{2}-1)) - m z \right] e^{ - i p_m \cdot x}, & M=\mu
\end{matrix}
\right. ,\\
\partial_M f =& \left\{ \begin{matrix}
- \frac{\sqrt{m}}{\sqrt{2 p_{m 0}}}  m z^{\frac{d}{2}} \sqrt{\frac{2}{\pi m z}} \cos \left[ \frac{\pi}{4} + \frac{\pi}{2} ((\Delta_\phi + 1) - \frac{d}{2}) - m z \right]   e^{-i p_m \cdot x}, & M=z \\
i \frac{\sqrt{m}}{\sqrt{2 p_{m 0}}} \left( \frac{p_{m \mu}}{m^2}\right)  m^2 z^{\frac{d}{2}} \sqrt{\frac{2}{\pi m z}} \cos \left[ \frac{\pi}{4} + \frac{\pi}{2} ((\Delta_\phi + 1) - (\frac{d}{2} - 1)) - m z \right] e^{-i p_m \cdot x}, & M=\mu
\end{matrix}
\right. .
\end{align} We find that $h_z = - \frac{1}{m} \partial f$ for $\Delta_\phi = \Delta_J - 1$, which is unsurprising since scalar and vector twists differ by 1.  We now only need to address the presence of a phase shift above.

The integral over the Lorentzian coordinates and $n$ can be performed in the large $m z$ limit in Eq. (\ref{ads_lsz_exchange}) to arrive at the following expression for the interacting piece of the matrix element: 
\begin{align}
S[A_z] &\propto \int_0^\infty d z_1 \;  z_1^{ \frac{d}{2}-2} \bigg\{ \left( m_1 m_2 + p_1 \cdot p_2 \right) \cos\left[ (m_1 - m_2) z_1 \right] \nonumber \\
& + \left( m_1 m_2 - p_1 \cdot p_2 \right) \sin \left[ (m_1 + m_2 ) z_1 - \left( \Delta_J - \frac{d}{2} \right) \pi \right] \bigg\} J_{\Delta - \frac{d}{2} } ( \sqrt{ s} z_1 ). \label{ads_lsz_exchange_int1}  
\end{align} For the second term in curly brackets, we can change the integration variable from $z_1$ to $y$ via $z_1 = y + \frac{1}{m_1 + m_2}  (\Delta_J - \frac{d}{2}) \pi$ and split the integral into a piece over the region $y \in [- \frac{1}{m_1 + m_2}  (\Delta_J - \frac{d}{2}) \pi, 0]$ and another over the region $y \in [0, \infty]$. The first region contributes negligibly since $J_{\Delta - \frac{d}{2}} ( \frac{\sqrt{s}}{m_1 + m_2}  (\Delta_J - \frac{d}{2}) \pi ) \sim \frac{1}{\Gamma( \Delta - \frac{d}{2} + 1)}$, and we are left with only the second region.  The second region is still negligible until $\sqrt{s} y + \frac{\sqrt{s}}{m_1 + m_2}  (\Delta_J - \frac{d}{2}) \pi \approx \sqrt{ \Delta}$, at which point $y \gg \frac{1}{m_1 + m_2}  (\Delta_J - \frac{d}{2}) \pi $, permitting us to drop the $\frac{\sqrt{s}}{m_1 + m_2}  (\Delta_J - \frac{d}{2}) \pi $ term.   We are then left with Eq. (\ref{ads_lsz_exchange_int1}) without the $(\Delta_J - \frac{d}{2}) \pi$ phase shift in the second term.  Since constant phase shifts are unimportant in the LSZ integral when $\Delta$ is very large, the difference between the phases in $\partial f$ and $h_z$ are irrelevant.

We can thus replace a single external gauge boson with a scalar when
\begin{equation}
\frac{\Delta_J }{m^2} \ll 1,
\end{equation} where the AdS curvature scale is set to unity. At this juncture, it is worthwhile to comment on our use of $m$ as a quantum number in our mode function expansion earlier. Since $p^2=m^2$, we can simply make the replacements $m \to |p|$ and allow $p_0 =\sqrt{\vec{p}^2 + m^2}$ to be the label for the external quantum states.  With this labeling, the above condition becomes
\begin{equation}
\frac{\Delta_J }{|p|^2} \ll 1,
\end{equation} 
and all gauge bosons may be replaced with scalars when \begin{align}
\frac{\Delta_J}{s} \ll 1,
\end{align} for a center of mass energy $s$. The AdS ET can then be stated as 
\begin{align} 
S[A_z]  =  \left[ - \sqrt{(\Delta_J - ( d - 1) ) (\Delta_J - 1)} \right]^N \left( \prod_{i=1}^N |p_i|^{-1} \right) S[\pi] \label{ads_et2}
\end{align} 
when 
\begin{align}
\frac{\Delta_J}{\Delta} \ll 1 \,\,\,\,\,\,\,\,\,\,\,\,\,\,\,\,\,\,\,\, \text{or} \,\,\,\,\,\,\,\,\,\,\,\,\,\,\,\,\,\,\,\, \frac{\Delta_J}{s} \ll 1. \label{large_exchange_limit}
\end{align}

Up to this point, the relevance of the matrix elements, $S$, in the language of AdS/CFT has been unclear.  Their physical significance is evident in the flat space limit (which is incidentally the relevant limit herein) as $S$-matrix elements (\cite{Fitzpatrick:2012cg}, \cite{Fitzpatrick:2011dm}), but it would be useful to understand them in the context of conformal correlators.  We note that we may express the correlator 
\begin{align}
\langle \mO (x) \dots \rangle =& \int d^d y dw \sqrt{g(w)} G_\partial( x- y; z) \Gamma(y, z; \dots),
\end{align}
where $G_\partial$ is the bulk-boundary propagator and $\Gamma$ is the usual bulk vertex function of interest here (\cite{Hamilton:2005ju}, \cite{wittenads}).  We may convolve the conformal correlator with some boundary source, $j$, to yield
\begin{align}
\int d^d x j(x; p) \langle \mO (x) \dots \rangle =& \int d^d x d^d y dw \sqrt{g(w)} j(x; p) G_\partial (x - y; z) \Gamma(y, z; \dots). \label{eqn:boundary_convolution}
\end{align}
We may choose $j(x;p)$ such that $\int d^d y j(x;p) G_\partial( x - y; z) = f(x, z; p)$ is a wave function used to define $S$.  For this to be the case, the left hand side of Eq. (\ref{eqn:boundary_convolution}) must be proportional to the Fourier transform of the conformal correlator to which $\Gamma$ is relevant.  For boundary currents, this proportionality factor must involve a projection of correlator onto a vector in the tangent bundle at the conformal boundary.  Consequently, Eq. (\ref{ads_et2}) can be interpreted as a statement about the relation between the Fourier transform of conformal correlators dual to the gauge and Goldstone fields.

\subsection{Implications for Correlation Functions of $J_\mu$}

At the end of \S \ref{exchange_op_sec}, we discussed how the fact that theories with sufficiently large masses of exchanged particles satisfy the ET at all external energy scales and thus open the possibility for a manifestation of the ET in correlation functions of fields, as opposed to $S$-matrix elements, to appear since such objects involve the sum of $S$-matrix-like objects over all energy scales. This should have strong implications for correlation functions of conformal current operators under the AdS/CFT program, which will be examined in this section.

The relationship between the divergence of the conformal current and the bulk gauge fields is established in Eq. (\ref{divergence_gauge_relation}), so the AdS ET should manifest through this scalar degree of freedom.  By defining 
\begin{equation}
\mathcal{O}_J \equiv (\Delta_J - (d -1) )^{-1} \partial \cdot J,
\end{equation} 
we may write the current as
 \begin{align}
J_\mu =& (\Delta_J - (d - 1)) \partial_\mu \partial^{-2} \mathcal{O}_J + J^{0}_\mu ,
\end{align} 
where 
\begin{align}
J^{0 a}_\mu \equiv& \left[ \delta^\nu_\mu - \partial_\mu \partial^{-2} \partial^\nu \right] J^a_\nu
\end{align}
 is the conserved part of the current. Expanding Eq. (\ref{divergence_gauge_relation}) in terms of creation/annihilation operators yields
 \begin{align}
\partial \cdot J &= (\Delta_J - (d -1)) \int d^{d-1} \vec{p} \int_0^\infty dm \, \left[  a_z f^\dagger_\partial + h.c.  \right]
\end{align} 
where  $f_\partial$ is the scalar wave function with the same scaling dimension as the gauge boson taken to the boundary, 
\begin{align}
f_\partial( \Delta_J, \vec{p}, m; x) =  \frac{1}{\Gamma(\Delta_J - \frac{d}{2} + 1)} \frac{ 1 }{ \sqrt{ p_{m 0}}} \left( \frac{m }{2} \right)^{\Delta_J - \frac{d-1}{2}}  e^{- i p_m \cdot x}. \label{boundary_scalar_wavefunction}
\end{align}  
As expected, $\partial \cdot J$ vanishes when $\Delta_J = d - 1$, corresponding to $m_A=0$.

While $\mathcal{O}^a_J$ is manifestly a scalar, it is a descendant in general theories since $J^a$ itself is primary.  However, when $\frac{\Delta_J}{\Delta} \ll 1$, Eq. (\ref{ads_et2}) shows that in correlation functions we may make the replacement 
\begin{align}
a_z \to - m^{-1} \sqrt{( \Delta_J - (d-1)) (\Delta_J -1)} b_\pi , \label{operator_replacement}
\end{align} 
where $b_\pi$ is the creation/annihilation operator for the corresponding Goldstone boson, which should have a primary dual. Defining the dual to the Goldstone as 
\begin{align}
\mathcal{O}_\pi &\equiv - \frac{1}{2  \left( \Delta_J - \frac{d}{2} + 1 \right)}  \sqrt{( \Delta_J - (d - 1)) (\Delta_J -1)} \int d^{d-1} \vec{p} \int_0^\infty dm \, \left[ b_\pi  f^\dagger_\partial (\Delta_\pi ) + h.c.  \right] \\
&= \int d^{d-1} \vec{p} \int_0^\infty dm \, \left[   a_z  f^\dagger_\partial (\Delta_\pi +1 ) + h.c.  \right],
\end{align} 

where the last line follows from Eqs. (\ref{boundary_scalar_wavefunction}) and (\ref{operator_replacement}),  we see $\mathcal{O}_\pi$ is the same as $\mathcal{O}_J$ for $\Delta_\pi = \Delta_J - 1$.  This is again unsurprising given the relative scaling of vectors and scalars to the boundary.

Recall that the non-interacting part of the matrix elements trivially satisfies the AdS ET and that the interacting part is insensitive to the scaling dimensions of the external particles when $\frac{\Delta_J }{\Delta} \ll 1$. The operators $\mathcal{O}_J$ and $\mathcal{O}_\pi$ may then be identified, and the expression of the AdS ET under the AdS/CFT prescription is thus 
\begin{align}
\mathcal{O}_J \to \mathcal{O}_\pi. \label{cft_et}
\end{align} 

Equivalently, we may state that $\mathcal{O}_J$ is approximately primary when computing correlation functions.   The usual techniques for computing correlators of primary operators for both $\mathcal{O}_\pi$ and $J^{0}_\mu$ may thus be used in theories in which $\frac{\Delta_J}{\Delta} \ll 1$.

To recapitulate what was shown from bulk AdS considerations:

\begin{itemize}
\item The $s=z$ wave function was shown to be analogous to the longitudinal polarization.
\item The AdS equivalent to large $|\vec{p}|$ in flat space was shown to be large $|p|$ or $s$.
\item For large interaction energies, $S[A_z]\propto S[\pi]$.
\item The manifestation of the AdS ET at the boundary is, essentially, the Ward identity: $\partial \cdot \langle J \dots \rangle \propto \langle \mO_\pi \dots \rangle$.
\end{itemize}

\section{Higher Spin AdS ET} \label{higher_spin}
We have shown in the previous sections that, in light of the Dyson-Schwinger equations, the ET results simply from the asymptotic equivalency of the the derivative of a scalar wave function and a vector wave function in the high-energy limit.  This implies an equivalence theorem relating spin-$l$ processes to lower spin processes can then be obtained by demonstrating that the spin-$l$ wave functions are asymptotically equivalent to symmetrized derivatives of lower-spin wave functions in the high-energy limit.  We construct such a theorem in this section.

To begin, we briefly review higher spin fields.  We wish to consider a massive real rank $l$ field, $\phi_{M_1 \dots M_l} (x,z)$, with scaling dimension $\Delta_J$ in an $AdS_{d+1}$ vacuum that is symmetric, traceless, and transverse:
\begin{align}
\phi_{M_1 M_2 \dots M_l} =& \phi_{M_2 M_1 \dots M_l} = \phi_{( M_i \leftrightarrow M_j )} \forall i,j\\
\nabla^{M_1} \phi_{M_1 \dots M_l} =& 0\\
{\phi^{M_1}}_{M_2 \dots M_l} =& 0.
\end{align}

Enforcing these conditions ensures that $\phi$ is an irreducible spin-$l$ representation of the AdS isometries, and is the spin-$l$ generalization of projecting out the unphysical divergence in gauge theories.  That is, the degrees of freedom that are projected out by these constraints should correspond to ``gauge" degrees of freedom.  Consequently, we may unambiguously take this field to the $CFT_d$ boundary in a ``gauge independent" way and identify the CFT dual in the Poincar\'{e} patch as
\begin{align}
\mO_{\mu_1 \dots \mu_l} (x) \equiv \lim_{z \to 0} \frac{1}{z^{\Delta_J - l}} \phi_{\mu_1 \dots \mu_l} (x,z). \label{eqn:cft_dual}
\end{align}

Per usual, we may expand the $\phi$ field in terms of mode functions.  Demanding that the mode functions be in the same representation of the AdS isometries as our field means the mode functions must satisfy the classical \textit{free} equations of motion for $\phi$\footnote{The (linearized) differential operator appearing in the free classical equations of motion is simply the Casimir of the AdS algebra, with the $\phi$ mass playing the role of the weight of the representation.},
\begin{align}
\left[ {\mathcal{D}^2}^{N_1 \dots N_l}_{M_1 \dots M_l} + (\Delta_J -l)(\Delta_J - (d-l)) \delta^{N_1}_{M_1} \dots \delta^{N_l}_{M_l} \right] (\phi_{free})_{N_1 \dots N_l} =0 \label{eqn:spinl-eom}
\end{align}
where $\mathcal{D}^2$ is a second order differential operator containing the Laplace-Beltrami operator, $\nabla^2$, as a linear contribution within it.  This suggests we can expect to parameterize the mode functions in the Poincar\'{e} patch using the momentum in the boundary coordinate directions.  Aditionally, a symmetric, traceless, transverse field of rank $l$ in $d+1$ dimensions that is generally massive has $D\equiv \left( \begin{matrix} d+l\\ l \end{matrix} \right) - 2$ degrees of freedom, which corresponds to $D$ polarizations.  Our expansion then takes the form
\begin{align}
\phi_{M_1 \dots M_l}(x, z) = \sum_{s=1}^D  \int \dbar^d p \; \left[ a_{s, p} (z)\, \phil_{M_1 \dots M_l} (s, p; x, z) + h.c. \right], \label{eqn:spinl-expansion}
\end{align}
with the notation $\dbar = \frac{1}{2\pi} d$.

Since we are parameterizing our mode functions using momentum, it is natural to foliate AdS in the $z$-direction and define a $z$-independent inner product over function space such that our mode functions are orthonormal\footnote{Previously, we foliated in time.  The definition of the inner product remains the same as that presented in appendix \S \ref{wave_function_review} with the replacement $t \to z$.},
\begin{equation}
\langle \phil(s,p) , \phil(s',p') = (2 \pi)^d \delta_{s, s'} \delta^d( p - p'). \label{eqn:mode_orthogonality}
\end{equation}

Then when our theory includes interactions, $a_{s,p}$ generally exhibits a dependence on $z$, which we have explicitly included in Eq. (\ref{eqn:spinl-expansion}).

Under the normalization Eq. (\ref{eqn:mode_orthogonality}), the $a$ algebra satisfies
\begin{align}
[a_{s,p}(z), a^\dagger_{s',p'}(z)] = (2 \pi)^d \delta_{s, s'} \delta^d(p -p'). \label{eqn:mode_algebra}
\end{align}

Since the $\phil$'s transform as simple representations of the AdS isometries, we are in a good position to consider the ET.

\subsection{AdS ET via Analysis of Wave Functions}
The ET is largely a statement about the degrees of freedom contained in a spin-$l$ field in its massless limit.  For a representation $V$ of a group $G$ to qualify as irreducible, the following must hold:
\begin{align}
G v =& V, \quad\quad \forall \, v \in V.
\end{align}

In flat space, the transversality constraint for a massive field implies all polarizations must be space-like. In other words, one can go to a frame in which there are only non-vanishing components for spatial indices.  In the massless limit, polarizations must be either space-like or light-like.  In particular, some linear combinations of the polarizations must now be light-like.  Light-like objects can only be boosted to other light-like objects, thus pulling the particular polarizations out of the spin-$l$ orbit into a spin-$(l-1)$ orbit:  $V_l \to V'_l \oplus V_{l-1}$, where $V'_l$ still transforms as a spin-$l$ representation, but is a smaller dimension than $V_l$ (it is spanned by only space-like polarizations).  In practice, this means for some $s$ that $\phil_{\mu_1 \dots \mu_l} (s,p)= \partial_{( \mu_1} \phill_{\mu_2 \dots \mu_l)}(p)$ in flat space. In AdS, this means we wish to find all $s$ such that
\begin{align}
\phil_{M_1 \dots M_l} (s,p) \underset{ \frac{\Delta_J}{|p|} \ll 1}{=}& \nabla_{( M_1} \phill_{M_2 \dots M_l)} (p) \nonumber\\
 \underset{ \frac{\Delta_J}{|p|} \ll 1}{=}& \partial_{( M_1} \phill_{M_2 \dots M_l)} (p),\label{eqn:spinl_et_statement}
\end{align} where $( \dots )$ denotes complete symmetrization of the indices.  The second line follows since the large $|p|$ limit, $\frac{\Delta_J}{|p|} \ll 1$, implies derivative terms dominate and the contribution of the Christoffel symbol is negligible. The $(l-1)$ mode function, $\phill$, may itself correspond to several potential polarizations, so we have foregone labeling it.

Now if $\phil$ and $\nabla \phill$ satisfy the same equations, then Eq. (\ref{eqn:spinl_et_statement}) holds.  In the large $|p|$ limit, we expect terms in Eq. (\ref{eqn:spinl-eom}) that go as $\partial^2 \equiv \eta^{\mu \nu} \partial_\mu \partial_\nu$ and $\partial_z^2$ to dominate; we additionally wish to maintain proper boundary asymptotics, so we insist on keeping terms that go as $\partial_z$ as well.  All terms that do not involve a derivative are to be ignored.  Consequently, Eq. (\ref{eqn:spinl-eom}) becomes 
\begin{align}
{\nabla^2}^{N_1 \dots N_l}_{M_1 \dots M_l} \phil_{N_1 \dots N_l} =0. 
\end{align}

Acting on $\phil$ with $g^{A B} \nabla_A \nabla_B$ and keeping only derivative terms yields
\begin{align}
\nabla^2_{scalar} \phil_{M_1 \dots M_l} -  2 g^{A B} \Gamma^C_{A (M_1} \partial_B \phil_{C \dots M_l)} =&0 \implies \nonumber \\
\left[ - z^{2} \partial_z^2  + (d-1) z \partial_z + z^2 \partial^2 \right] \phil_{M_1 \dots M_l} - 2 g^{A B} \Gamma^C_{A (M_1} \partial_B \phil_{C \dots M_l)} =&0, \label{eqn:spinl_eom_pre_limit}
\end{align}
where the symmetrizer $( \dots )$ acts on the $M$ indices only. To be explicit about the difference between the scalar equation of motion and the spin-$l$ equations, note that each index contributes another term of the form $\Gamma \partial \phil$.  Explicitly, this operator is
\begin{align}
\Gamma^C_{A M} g^{A B} \partial_B = z \left[ \delta^C_M \partial_z - \delta^C_z \partial_M \right], \label{eqn:index_contribution}
\end{align}
where the term $-\delta^z_M \eta^{B C} \partial_B$ since its action is trivial on transverse fields in this limit.  The first term in Eq. (\ref{eqn:index_contribution}) simply differentiates $\phil$ with respect to $z$ and adds $l$ of such terms.  This can be assimilated into the operator in brackets in Eq. (\ref{eqn:spinl_eom_pre_limit}) to send $(d-1) \to (d - 2 l -1)$.  The second term in Eq. (\ref{eqn:index_contribution}) simply results in the unsurprising symmetrized derivative term $z \partial_{(M_1} \phil_{z \dots M_l)}$.  Revisiting Eq. (\ref{eqn:spinl_eom_pre_limit}) under this prescription yields
\begin{align}
\left[ - z^{2} \partial_z^2  + (d - 2l -1) z \partial_z + z^2 \partial^2 \right] \phil_{M_1 \dots M_l} + 2z \partial_{(M_1} \phil_{z \dots M_l)}=&0. \label{eqn:spinl_eom_limit}
\end{align}

In flat space, the light-like polarization in the massless limit is special because its orbit under the Lorentz group is just other light-like polarizations.  Recall in the spin-1 case in AdS that mode functions with a non-vanishing $z$-component could not be transformed to mode functions with vanishing $z$-component when acted upon by the AdS isometry group in the massless limit.  It was, indeed, the analogue to the light-like longitudinal polarization as a consequence.  We keep this fact in mind and note the special appearance of $\phil_{z \dots M_l}$ in Eq. (\ref{eqn:spinl_eom_limit}).   Proceeding, we choose our mode functions such that they divide naturally into those with no $z$-component (for a single index) and those with only the $z$-component as the degree of freedom (with non-$z$-components appearing as derivatives of the $z$-component) and consider the equations of motion for $M_1=z$:
\begin{align}
\left[ - z^{2} \partial_z^2  + (d - 2(l - 1) -1) z \partial_z + z^2 \partial^2 \right] \phil_{z M_2 \dots M_l} + 2z \partial_{(M_2} \phil_{z z \dots M_l)}=&0. \label{eqn:spinl_eom_limit_z}
\end{align}
This is precisely the equations $\phill$ satisfies. Differentiating Eq. (\ref{eqn:spinl_eom_limit_z}) with respect to $x^{M_1}$, symmetrizing, and throwing out terms without derivatives yields
\begin{align}
\left[ - z^{2} \partial_z^2  + (d - 2l -1) z \partial_z + z^2 \partial^2 \right] \partial_{( M_1} \phill_{ M_2 \dots M_l)} + 2z \partial_{(M_1} \partial_{(M_2} \phill_{ \dots M_l))}=&0.
\end{align}
This is exactly Eq. (\ref{eqn:spinl_eom_limit}).  We thus conclude that mode functions whose only degrees of freedom come from setting one of its indices to $z$ are exactly the mode functions we seek in the ET.  That is, mode functions such that
\begin{align}
\phil(s=z, p)_{M_1 \dots M_l} = \left\{
\begin{matrix}
\phil(s=z, p)_{z \dots M_l} &M_1=z\\
\partial_{(\mu} \partial^{-2} \partial_z \phil(s=z, p)_{z \dots M_l)} & M_1 = \mu 
\end{matrix}
\right. .
\end{align}

We may thus write our original AdS field in the ``high-energy" limit as
\begin{align}
\phi_{M_1 \dots M_l} \to \phi^0_{M_1 \dots M_l} + \nabla_{( M_1} \theta_{M_2 \dots M_l)}
\end{align}
for some spin-$(l-1)$ field $\theta_{M_1 \dots M_{l-1}}$. 

If the trace and non-transverse components were left as gauge degrees of freedom, the above would describe a theory for which the Lagrangian was of the form $\mathcal{L}(\phi_{M_1 \dots M_l} - \nabla_{( M_1} \theta_{M_2 \dots M_l)} )$.  The spontaneously broken gauge transformations are given by $\delta \phi_{M_1 \dots M_l} = \nabla_{( M_1} \epsilon_{M_2 \dots M_l)}$ while the Goldstone mode $\theta$ is simply shifted by $\epsilon$.

It is worthwhile to remark that we could have repeated this process of setting an index to $z$ in Eq. (\ref{eqn:spinl_eom_limit}) to obtain a tower in which we ultimately conclude
\begin{align}
\phil_{\text{scalar}}(p)_{M_1 \dots M_l} = \partial_{(M_1} \dots \partial_{M_l)} \varphi(p)
\end{align}
for a scalar mode, $\varphi$.  This mode function contributes the most in the high-energy limit, and thus, unsurprisingly, we can write the dominant contribution to the theory as
\begin{align}
\phi_{M_1 \dots M_l} \to \nabla_{( M_1} \dots \nabla_{M_l)} \theta.
\end{align}

\subsection{Spin-l AdS ET at the Boundary}
As in the spin-$1$ case, a free theory trivially satisfies the ET.  If the interactions that are added are only excited at high-energies (short distances), corresponding to heavy exchange operators, $\Delta \gg \Delta_J$, then the equivalence theorem holds in position space when all scales of $|p|$ are integrated over.  At the boundary, the equivalence theorem then takes the form
\begin{align}
\partial^{\mu} \mO_{\mu \mu_1 \dots \mu_{l-1}} =& \lim_{z\to 0} \frac{1}{z^{\delta - l}} z^{d-1} \partial_z \left[ \frac{1}{z^{d-1}} \phi_{z \mu_1 \dots \mu_{l-1}} \right] \nonumber \\
=&  \lim_{z\to 0} \frac{1}{z^{\delta - (l-1)}} \phi_{ \mu_1 \dots \mu_{l-1}}  \nonumber \\
=& \mO_{\mu_1 \dots \mu_{l-1}},
\end{align}
which is to say the spin-$l$ conformal current is not conserved and its divergence is approximately a primary spin-$(l-1)$ current when computing correlators.  Continuing with the tower prescription discussed in the previous section, we may write
\begin{align}
\mO_{\mu_1 \dots \mu_l} = \mO^0_{\mu_1 \dots \mu_l} + \partial_{(\mu_l} \partial^{-2} \mO^0_{\mu_1 \dots \mu_{l-1})} + \dots + \partial_{\mu_1} \dots \partial_{\mu_l} \left( \partial^{-2} \right)^l \mO,
\end{align}
where each $\mO^0$ is a conserved primary current.

Of course, this allows one to write approximately
\begin{equation}
\mO_{\mu_1 \dots \mu_l} \approx  \partial_{\mu_1} \dots \partial_{\mu_l} \left( \partial^{-2} \right)^l \mO,
\end{equation}
thus reducing the problem of computing spin-$l$ conformal correlators to computing computing a scalar correlator.

\subsection{Generalization to CFT ET}

That the divergence of a non-conserved conformal current is approximately primary in theories with particular bulk couplings is the most interesting consequence of the AdS ET.  While bulk gauge theories are always dual to theories with conformal currents, the inverse mapping is not unique, and it is of interest to confirm that the divergence of conformal currents is generally primary in particular limits under different bulk theories.  It would also be useful to examine any additional consequences or constraints of the ET for conformal theories.  We thus seek a purely conformal, bottom-up approach to the ET.

The wave functions that played such a central role in the purely AdS approach to the ET are irreducible representations of the conformal group.  The matrix elements, $S$, are built out of products of these functions and, consequently, can be expressed as sums over other irreducible representations of the conformal group. This is akin to the procedure of expanding conformal correlators in irreducible representations of the conformal group as conformal blocks.  Since we may interpret the matrix elements as Fourier transforms of conformal correlators, and the AdS ET arises from a relationship between scalar and vector wave functions, it seems natural to examine the ET on the CFT side as a relationship between conformal blocks of currents and of scalars. In the following sections, we thus analyze the CFT ET using the machinery of the conformal block expansion.

\section{Equivalence Theorem in CFTs\label{sec:cftet}} In this section, we will generalize the ET in terms of conformal blocks. We begin with a warm-up example in two dimensions, which  will serve as both a simple introduction and a distinct contrast to the more interesting four dimensional case. We emphasize that the $d=2$ example is merely an explication of ideas that are known in the literature (see, e.g. \cite{ginsparg}, \cite{osborn}, \cite{recursion}, and \cite{Belavin:1984vu}). It is well known that in $d=2$ the conformal group naturally breaks up into holomorphic/antiholomorphic (also called `left moving/right moving') parts. Consequently, the conformal blocks themselves factorize into holomorphic and antiholomorphic terms, which greatly facilitates the analysis of spinning correlators. We will then move onto the more involved $d=4$ case, where we will review the index-free formalism of \cite{spin} and \cite{blocks} and determine the ET as a statement of a spin-1 CFT current.

\subsection{Two Dimensional Warm-Up\label{sec:2d}} A hallmark feature of 2-$d$ CFTs is that one may change coordinates to the complex plane by noting that the line element may be expressed as $ds^2 = dx^2 + dy^2 = dzd\zbar$, where $z = x + iy$. Therefore, we can write the metric as \be g_{ab} = \begin{pmatrix} 0 & \frac{1}{2}\\ \frac{1}{2} & 0 \end{pmatrix},\ee where $a$ and $b$ denote $z$ or $\zbar$. Under this coordinate change, the current $J^\mu$ can be mapped to a two component vector \be \begin{pmatrix} 
  J_z \\
  J_{\zbar}
  \end{pmatrix} = \begin{pmatrix}J_x + i J_y \\
  J_x - i J_y
\end{pmatrix}.\ee If one is interested in correlators that involve only $J$ and hermitian operators, then it is clear that one may consider only one component and obtain the other by complex conjugation. In general, $J$ may be a higher spin current, in which case one must consider the total independent degrees of freedom classified by their weights $(h,\hba)$. We will elaborate on this point when we examine the spin-3 current.

Unitarity bounds of CFT currents, which are typically calculated by bounding the norm of descendant states, have clear bulk interpretations. If the scaling dimension of the spin-$\ell$ current exceeds $d-2+\ell$, then this corresponds to a massive bulk gauge boson. Conversely, when the bound is saturated, the current must be dual to massless bulk gauge boson and $\partial \cdot J = 0$ so that no degrees of freedom are lost or gained. It is evident then that any statement of the equivalence theorem in a traditional CFT must involve only those currents that do not saturate the bound, so that there are longitudinal bulk propagating degrees of freedom. So  before we move onto the ET, it will be worthwhile to clarify how these bounds show up in the divergences of correlation functions. In two dimensions, it is possible to check this bound by considering the divergence around a small ball $\Omega$ \begin{align} \int_\Omega d^2 x \partial_\mu K^\mu, \end{align} where $K^\mu$ may be any correlator of $J^\mu$. We can write this as a surface integral \begin{align} \int_\Omega d^2 x \partial_\mu K^\mu &= \int_{\partial \Omega} d A_{\mu} K^{\mu} \nonumber \\
&=\frac{i}{2}\int_{\partial \Omega} \left(d \zbar K^{z} - dz K^{\zbar} \right), \label{eqn:div} \end{align} where $dA_{\mu}$ runs along the counterclockwise contour. For a given correlator involving an arbitrary number of operators, the above integrals are difficult to evaluate in full generality. However, we may consider a simple example of a three point function involving only a single spin-1 current \be \langle J^z(z_1,\zbar_1) \mO_2 (z_2,\zbar_2) \mO_3(z_3,\zbar_3)\rangle = \frac{1}{|z|^{2h_{23} + \Delta_J} |z-1|^{2h_{32} + \Delta_J}} \left[\frac{\zbar(\zbar-1)}{z(z-1)} \right]^{1/2}, \ee where $h_{ij} \equiv h_i - h_j$ and we have mapped $z_1 \to z$, $z_2 \to 0$, $z_3 \to 1$. The $\bar{z}$ component is identical, except the exponent of the square bracket term is $-1/2$. It is easy to check that Eq. (\ref{eqn:div}) vanishes up to contact terms when $\Delta = \ell = 1$ and when the scaling dimensions of the two scalars are the same. Although all of the above discussion is rather obvious, it is worth mentioning because the expressions we will soon encounter (which are divergences of correlators) may not appear to be zero up to contact terms at first glance when $\Delta_J= \ell$, but internal consistency can be checked using the above method.

\subsubsection{Spin-1 Current} 
Consider, then, the four-point function \be G_4 \equiv \langle J_1(z_1,\zbar_1) \mO(z_2,\zbar_2) \mO(z_3,\zbar_3) \mO(z_4,\zbar_4) \rangle, \ee where we will denote $J^z \equiv J$ and $J^{\zbar} \equiv \jbar$. For computational simplicity, we will take all three scalars to have the same scaling dimensions so \begin{align}  h_{\mO_i} = \hba_{\mO_i} \equiv h = \frac{\Delta}{2}, \,\,\,\,\,\,\,\,\,\,\,\,\,\,\,\,\,\,\,\,\,\,\,\,\,\,\, (i = 1,2,3). \label{c}\end{align} It is important to distinguish the spin of AdS$_3$ gauge boson and the spin of the current. While it is true that $J^{\mu}$ is a spin-1 current in that it has one index, the reducibility of the conformal group lets us classify correlator purely by the \textit{components} of $J^a$, as long as $J$ is primary\footnote{Otherwise a component of a tensor with $a$ $z$ indices and $b$ $\zbar$ indices would not transform as \be J^{z_1 z_2 \dots z_a \zbar_1 \zbar_2 \dots \zbar_b}(z,\zbar) \to \left(\pader{f}{z} \right)^a \left(\pader{\bar{f}}{\zbar} \right)^b J^{z_1 z_2 \dots z_a \zbar_1 \zbar_2 \dots \zbar_b}(z,\zbar), \ee with $z\to f(z)$.}. Here, the two spin states, classified by $(h,\hba) = (1,0)$ and $(0,1)$, correspond to $J$ and $\jbar$, respectively. The correlation function can be written as an overall scale term times a general function of the conformally invariant cross ratio $\eta \equiv \frac{z_{12} z_{34}}{z_{13} z_{24}}$: \begin{align} G_4^z &= \frac{1}{|z_{12}|^{\Delta_j+\Delta} } \frac{1}{|z_{34}|^{2\Delta}} \left(\frac{|z_{24}|}{|z_{14}|} \right)^{\Delta_j-\Delta}\left(\frac{\zbar_{12} z_{24} \zbar_{14}}{z_{12} \zbar_{24} z_{14}} \right)^{1/2} f_1(\eta,\ebar) \\ 
G_4^{\zbar} &= \frac{1}{|z_{12}|^{\Delta_j+\Delta} } \frac{1}{|z_{34}|^{2\Delta}} \left(\frac{|z_{24}|}{|z_{14}|} \right)^{\Delta_j-\Delta}\left(\frac{\zbar_{12} z_{24} \zbar_{14}}{z_{12} \zbar_{24} z_{14}} \right)^{-1/2} f_2(\eta,\ebar) \end{align} How should one deal with the functions $f_1$ and $f_2$? The four point function may be regarded as gluing together three point functions via the insertion of states corresponding to exchanged operators of weights $(h_e,\hba_e)$\footnote{Note here that the divergence of the correlator is given by \begin{align}\p_\mu G_4^\mu = \pbar G_4^z + \p G_4^{\zbar} &=\sum_{h_e,\hba_e} \left[\pbar \langle J \mO | h_e, \hba_e \rangle \langle h_e, \hba_e | \mO \mO \rangle + \p \langle \jbar \mO | h_e, \hba_e \rangle \langle h_e, \hba_e | \mO \mO \rangle  \right] \nonumber \\
&=\sum_{h_e,\hba_e} \left[ \pbar \langle J \mO \mO_{h_e,\hba_e} \rangle + \p \langle \jbar \mO \mO_{h_e,\hba_e} \rangle \right]\langle h_e, \hba_e | \mO \mO \rangle. \label{threept} \end{align} The term in the square brackets is $\p_\mu \langle J^\mu \mO \mO_{h_e, \hba_e} \rangle$. Therefore, if the three point functions involving $J$ and $\jbar$ satisfy the Ward identity, it is evident that the four point function will as well.}: \be G_4 =  \sum_{h_e,\hba_e}\langle J_1 \mO | h_e, \hba_e \rangle \langle h_e, \hba_e | \mO \mO \rangle.\ee  One can then package the contribution of a given primary and descendants to the four point function in the form of ``conformal blocks". That is, if the $f_i$ admit the following expansion: \be f_i(z,\zbar) \sim \sum_{h_e, \hba_e} \lambda_{h_e,\hba_e}\left[ \mathcal{L}_{h_e,\hba_e}(z,\zbar) + \mathcal{R}_{\hba_e, h_e}(z,\zbar) \right],\ee then $\mathcal{L}_{h_e,\hba_e}$ represents the contribution of a given primary and its descendants to the four point function and the second term $\mathcal{R}_{\hba_e,h_e}$ represents the contribution of its conjugate, and the coefficients $\lambda_{h_e,\hba_e}$ characterize the dynamics of the bulk theory. It is important to emphasize that the sum above runs only over the weights of exchanged primaries. The contributions of a single primary and its descendants have been summed into the blocks $\mathcal{L}$ and $\mathcal{R}$, whereas in Eq. (\ref{threept}), the sum runs over the weights of all exchanged operators. If all the exchanged operators correspond to scalar primaries and descendants (as we have assumed thus far), then $\mathcal{R}$ should be obtainable by simply swapping the arguments of $\mathcal{L}$. Consistent with the theme of left/right classification in two dimensional CFTs, we see that then the expansion would factorize in a trivial way.

Dolan and Osborn (\cite{dolan}) determined the functions $\mathcal{L}_{h_e,\hba_e}$ and $\mathcal{R}_{\hba_e,h_e}$ in terms of hypergeometric functions when the exchanged operators were symmetric traceless primaries. As foreseen, these functions factorize into left and right parts \begin{align} \mathcal{L}_{h_e,\hba_e} = k_{2h_e}(\eta) k_{2\hba_e}(\ebar) &= \eta^{h_e}\, {_2 F_1}(h_e - h_j + h, h_e; 2 h_e; \eta) \nonumber \\
&\times \ebar^{\hba_e}\, {_2 F_1}(\hba_e-h_j + h + 1,\hba_e;2\hba_e;\ebar), 
\end{align} and \begin{align} \mathcal{R}_{\hba_e,h_e} =  \mathcal{L}_{h_e,\hba_e}(\eta \leftrightarrow \ebar).\end{align} Therefore, the four point function is given by \begin{align} G_4^{a} &\sim S^a \sum_{h_e,\hba_e} \lambda_{h_e,\hba_e}^a \left(\eta^{h_e} \, {_2 F_1}(h_e - h_j + h, h_e; 2 h_e; \eta) \times \ebar^{\hba_e}\, {_2 F_1}(\hba_e-h_j + h + 1,\hba_e;2\hba_e;\ebar) + \eta \leftrightarrow \ebar  \right) \nonumber \\
&\equiv S^a \sum_{h_e,\hba_e} \lambda_{h_e,\hba_e}^a \left[k_{2h_e}(\eta) k_{2\hba_e}(\ebar) + k_{2h_e}(\ebar) k_{2\hba_e}(\eta) \right], \end{align} where we have encapsulated the scale term (the prefactors of the $f_i$) as $S^a$ (or $S$ and $\bar{S}$ for brevity). It is important to mention that since there are two degrees of freedom corresponding to the two helicity states of the bulk gauge boson, $f_1$ and $f_2$ (which we will call $f$ and $\bar{f}$ since they are associated with the $z$ and $\zbar$ components, respectively) will have independent expansion coefficients. It is clear that the divergence $\pbar G + \p \bar{G}$ will clearly result in the divergence of the scale terms ($\pbar S f + \p \bar{S} \bar{f}$) plus the divergence of the blocks themselves ($S \pbar f + \bar{S} \p \bar{f}$). It is the latter that will be of importance to us since the divergence of the scale term will result in another scale term, but the blocks themselves will remain unchanged. Let us consider the term $\bar{S} \p \bar{f}$. Apart from the fact that the OPE coefficients are different, there is no difference between $f$ and $\bar{f}$, so the analysis that will follow applies to the term $S \pbar f$ as well. Lastly, the blocks are functions of the anharmonic ratio and its conjugate, $\eta$ and $\ebar$ while the derivative acts with respect to $z$ or $\zbar$. While it is true that we may fix coordinates such that $\eta \to z$ and $\ebar \to \zbar$, there is no unique map that does this without causing the scale term to diverge. We must then take $\p = \p \eta \pader{}{\eta}$ and $\pbar = \pbar \ebar \pader{}{\ebar}$. With this out of the way, we have\begin{align} \p \bar{f} &= \sum_{h_e,\hba_e}\lambda_{h_e,\hba_e}^{\zbar} \bigg[ \p \eta \bigg(h_e \eta^{h_e-1} {_2 F_1}(h_e - h_j + h, h_e; 2 h_e; \eta) \nonumber \\
& + \frac{1}{2}(h+h_e-h_j) \eta^{h_e}{_2 F_1}(1+h+h_e-h_j, h_e +1,; 1+2 h_e; \eta)  \bigg) k_{2\hba_e}(\ebar) \nonumber \\
&+\p \eta \bigg( \hba_e \eta^{\hba_e -1} {_2 F_1}(\hba_e - h_j + h + 1, \hba_e; 2 \hba_e; \eta) \nonumber \\
&+ \frac{1}{2}(h+\hba_e-h_j+1) {_2 F_1}(2+h+\hba_e-h_j,\hba_e +1; 1+2 \hba_e; \eta)  \bigg)k_{2h_e}(\ebar) \bigg]. \end{align} Already at this level, we can see a very simple version of the ET. Suppose $h_e$ is large compared to both $h_j$, $h$, and is comparable to $\hba_e$. The second condition means that $\hba_e = h_e - \ell \approx h_e$ i.e. the twists of the exchanged operators are dominated by their scaling dimensions\footnote{More precisely, the exchanged operators may be classified by their irreducible components, given by weights $(h_e, h_e-\ell)$, $(h_e-1,h_e-\ell+1)$, etc. down to $(h_e-\ell, h_e)$. If the twists of these components are not dominated by $\ell$, i.e. $\ell \ll h_e$, then we may assume that $\hba_e \sim h_e$ for all states in the $h_e \to \infty$ limit.}. Then, the above result readily factorizes into scalar blocks. This can be seen by the following trivial expansions: \begin{align} &\eta^{h_e-1} \equiv \eta^{h_e(1-\eps)} \approx \eta^{h_e} + \mO(\eps); \\
&{_2 F_1}(1+h+h_e-h_j, h_e + 1; 1+2h_e;\eta) \approx {_2 F_1}(h_e(1 + \eps), h_e(1 + \eps); (\eps+2)h_e;\eta) \nonumber \\
&=1 + \frac{(1+\eps)^2 h_e \eta}{2+\eps} + \frac{(1+\eps)^2 h_e (1+h_e(1+\eps))^2\eta^2}{2(2+\eps)(1+h_e(2+\eps))} + \dots \nonumber \\
&= {_2 F_1}(h_e, h_e; 2h_e;\eta) + \mO(\eps).
 \end{align} Plugging these relations, in the $h_e \to \infty$ limit, we see that \begin{align} \p \bar{f} &\to \sum_{h_e,\hba_e} \lambda_{h_e,\hba_e}^{\zbar} \p \eta\left[ \left( h_e + \frac{1}{2}(h+h_e-h_j) \right) k_{2h_e}(\eta) k_{2\hba_e}(\ebar) + \left( \hba_e + \frac{1}{2}(h+\hba_e-h_j+1) \right)k_{2\hba_e}(\eta) k_{2h_e}(\ebar) \right] \nonumber \\
 &\approx \sum_{h_e,\hba_e} \lambda_{h_e,\hba_e}^{\zbar} \p \eta \left( h_e + \frac{1}{2}(h+h_e-h_j) \right) \left[k_{2h_e}(\eta) k_{2\hba_e}(\ebar) + k_{2\hba_e}(\eta) k_{2h_e}(\ebar) \right], \label{eqn:divblock}
 \end{align} where again we have assumed that $\hba_e \sim h_e$. We immediately see in the above equation the appearance of the scalar blocks of the form $k_i (\eta) k_j(\ebar) + k_j(\eta) k_i(\ebar)$, weighted by some coefficient that depends on the twist of the exchanged operator. Here, we see that the divergence leads to a term that is at most linear in the twist whereas in the four dimensional case, we will see that there exists a tower of operators beginning with terms proportional to $\Delta^{s+1}$, where $s$ is the spin of the current, down to $\mO(1)$ terms. By contrast, we see that in two dimensions, all terms contribute equally in the power of the exchanged weights. It should also be evident that the other terms in the divergence do not spoil the analysis above (we will compute all such terms in the four dimensional case and show this in full generality, as this is just a pedagogical example). The divergence of the scale terms will result in other scale terms, but they remain prefactors of scalar blocks. The remaining term $\pbar f$ will result in Eq. (\ref{eqn:divblock}) with the replacement $\lambda_{h_e,\hba_e}^{\zbar} \to \lambda_{h_e,\hba_e}^{z}$ and $\p \eta \to \pbar \ebar$.


\subsubsection{Spin-3 Current} Not surprisingly, higher spin correlators in two dimensional CFTs are as easy to handle as the spin-1 case and the ET for these objects essentially reduces to the analysis of last section. Consider \be G_4 \equiv \langle S(z_1) \mO(z_2) \mO(z_3) \mO(z_4) \rangle, \ee where $S$ is a spin-$3$ operator. We must clarify how many independent degrees of freedom there are. After the usual change of coordinates, where $S^{\mu\nu\sigma} \to S^{abc}$, we may eliminate 4 degrees of freedom assuming $S$ is symmetric and therefore components of $S$ are discriminated only by the number of $z$ and $\zbar$ indices\footnote{Unitarity places bounds on the weights $(h,\hba)$. However, there is no loss in generality if we relax these bounds.}. They are \be S^{zzz}, \,\,\,\,\,\,\,\,\,\,\,\, S^{zz\zbar}, \,\,\,\,\,\,\,\,\,\,\,\, S^{z\zbar\zbar}, \,\,\,\,\,\,\,\,\,\,\,\, S^{\zbar\zbar\zbar}.\ee  The idea is each of the above components are themselves irreducible components of the global conformal group and we must compute the associated four point function for each of them. Furthermore, we would like to emphasize that since these really are independent components with different weights, we will have four distinct functions of $\eta$ and $\ebar$, which we will label by \be f_1(\eta,\ebar), \,\,\,\,\,\,\,\,\,\,\,\,\,\,\,\, f_2(\eta,\ebar), \,\,\,\,\,\,\,\,\,\,\,\,\,\,\,\, f_3(\eta,\ebar), \,\,\,\,\,\,\,\,\,\,\,\,\,\,\,\, f_4(\eta,\ebar), \ee analogous to the spin-1 case.  The spin-3 current may then be regarded as four independent expansions in conformal blocks. The following expressions for the components, labeled by the operator weights, are then evident: \begin{align} (3,0):\,\,\,\,\,\,\,\, G_4^{zzz} &= \frac{1}{|z_{12}|^{\Delta_j+\Delta} } \frac{1}{|z_{34}|^{2\Delta}} \left(\frac{|z_{24}|}{|z_{14}|} \right)^{\Delta_j-\Delta}\left(\frac{\zbar_{12} z_{24} \zbar_{14}}{z_{12} \zbar_{24} z_{14}} \right)^{3/2} f_1(\eta,\ebar) \label{eqn:first} \\
(2,1):\,\,\,\,\,\,\,\, G_4^{zz\zbar} &= \frac{1}{|z_{12}|^{\Delta_j+\Delta} } \frac{1}{|z_{34}|^{2\Delta}} \left(\frac{|z_{24}|}{|z_{14}|} \right)^{\Delta_j-\Delta}\left(\frac{\zbar_{12} z_{24} \zbar_{14}}{z_{12} \zbar_{24} z_{14}} \right)^{1/2} f_2(\eta,\ebar) \\
(1,2):\,\,\,\,\,\,\,\, G_4^{z\zbar\zbar} &= \frac{1}{|z_{12}|^{\Delta_j+\Delta} } \frac{1}{|z_{34}|^{2\Delta}} \left(\frac{|z_{24}|}{|z_{14}|} \right)^{\Delta_j-\Delta}\left(\frac{\zbar_{12} z_{24} \zbar_{14}}{z_{12} \zbar_{24} z_{14}} \right)^{-1/2} f_3(\eta,\ebar) \\
(0,3):\,\,\,\,\,\,\,\,G_4^{\zbar\zbar\zbar} &= \frac{1}{|z_{12}|^{\Delta_j+\Delta} } \frac{1}{|z_{34}|^{2\Delta}} \left(\frac{|z_{24}|}{|z_{14}|} \right)^{\Delta_j-\Delta}\left(\frac{\zbar_{12} z_{24} \zbar_{14}}{z_{12} \zbar_{24} z_{14}} \right)^{-3/2} f_4(\eta,\ebar).\label{eqn:last}\end{align} The divergence can then be expressed as the two by two matrix \be \p_\mu G_4^{\mu\nu\sigma} \to \begin{pmatrix} \pbar G_4^{zzz} + \p G_4^{\zbar zz} & \pbar G_4^{zz\zbar} + \p G_4^{\zbar z\zbar}\\ \pbar G_4^{z\zbar z} + \p G_4^{\zbar \zbar z} & \pbar G_4^{z\zbar\zbar} + \p G_4^{\zbar \zbar \zbar} \end{pmatrix}. \label{eqn:divmatrix} \ee Before we go any further, let us take stock of some recurring themes. Four point functions of any spin current with three other scalars can be characterized completely by the exponent of the $\frac{\zbar_{12} z_{24} \zbar_{14}}{z_{12} \zbar_{24} z_{14}} \equiv \beta $ term. If we denote the common prefactor in the above Eqs. (\ref{eqn:first}) - (\ref{eqn:last}) as $\alpha$, then they reduce simply to the form $\alpha \beta^k f_i(\eta,\ebar)$. For a spin-3 current, have 3+1 degrees of freedom and so the irreducible states are distinguishable only by their values of $k$ given above: $3/2$ , $1/2$, $-1/2$, and $-3/2$ (corresponding to $s=3$, $1$, $-1$, and $-3$) along with their partial wave expansion coefficients. The general pattern we see is that components of symmetric four point functions involving a higher spin current $\langle J^s \mO \mO \mO\rangle $ are distinguishable by $k=\frac{s}{2}, \frac{s-2}{2}, \dots, -\frac{s}{2}$, corresponding to the $zzz\dots z$, $zzz\dots \zbar$, $\dots$, $\zbar\zbar\zbar \dots \zbar$ components and each independent component will be associated with a different function of $\eta$ and $\ebar$ (that is, we have as many independent functions as independent components). Therefore, it is simple to extend the ET to hold for any higher spin correlator. With all this in mind, Eq. (\ref{eqn:divmatrix}) can be written as a ``scale" matrix (which includes the divergence of the scale terms) plus the derivatives acting on the $f_i$: \begin{align} &\begin{pmatrix} f_1\left(\pbar \alpha \beta^{3/2} + \frac{3\alpha \beta^{1/2} \pbar \beta}{2} \right) + f_2\left(\p \alpha \beta^{1/2} + \frac{ \alpha \beta^{-1/2}\p \beta}{2} \right) & f_1\left(\pbar \alpha + \frac{ \alpha \beta^{-1/2} \pbar \beta}{2} \right) + f_3\left(\p \alpha \beta^{-1/2} - \frac{\beta^{-3/2} \alpha \p \beta}{2} \right)\\ f_1\left(\pbar \alpha + \frac{\alpha \beta^{-1/2} \pbar \beta}{2} \right) + f_3\left(\p \alpha \beta^{-1/2} - \frac{\beta^{-3/2} \alpha \p \beta}{2} \right) & f_2\left(\pbar \alpha \beta^{-1/2} - \frac{\beta^{-3/2}\alpha \pbar \beta}{2} \right) + f_4\left(\p \alpha \beta^{-3/2} - \frac{3 \alpha \beta^{-5/2} \p \beta}{2} \right) \end{pmatrix} \nonumber \\
&+ \alpha \begin{pmatrix} \beta^{1/2}\p f_2 + \beta^{3/2}\pbar f_1 &  \beta^{1/2} \pbar f_1 +  \beta^{-1/2} \p f_3 \\  \beta^{1/2} \pbar f_1 +  \beta^{-1/2} \p f_3 &  \beta^{-1/2} \pbar f_3 +  \beta^{-3/2} \p f_4 \end{pmatrix}. \label{eqn:spin3div}  \end{align} Clearly, the first term leaves the conformal blocks unchanged. Combined with the result of the last section, \be \p f_i \approx  \sum_{h_e,\hba_e} \lambda_{h_e,\hba_e}^{i} \p \eta \left( h_e + \frac{1}{2}(h+h_e-h_j) \right) \left[k_{2h_e}(\eta) k_{2\hba_e}(\ebar) + k_{2\hba_e}(\eta) k_{2h_e}(\ebar) \right], \ee we see that the second term of Eq. (\ref{eqn:spin3div}) also reduces to a matrix whose components are scalar blocks in the high energy limit of exchanged operator weights.


\subsection{Higher Dimensions\label{sec:4d}} We will now move onto the more interesting case of the ET in $d=4$. The four dimensional case is qualitatively different from the two dimensional case (i.e. the conformal group does not factorize) and analysis of correlation functions involving spinning fields is often a tremendous computational exercise. The required computations in position space are (in principle) tractable but in practice, the propagation of indices and counting of possible tensor structures makes for a difficult calculation.

Recently, the authors of \cite{spin} developed an index-free formalism where the index of a current $J$ is encoded into auxiliary vectors $z_{\mu_i}$. Great simplifications occur when one lifts the index-free correlator to embedding space, where the $x$'s project to $P$'s and the $z$'s project to $Z$'s. Namely, we find that terms which are $\mO(P_i\cdot Z_i)$ and $\mO(P^2, Z^2)$ are redundant and so we do not need to include them in the calculation. Although this method facilitates a great deal of intermediate calculations, we will have to project back down to physical space at the end of the day in order to consider the conservation operator. For completeness, we will review this formalism below, but readers familiar with these concepts can skip to \S \ref{sec:spin1current}.

\subsubsection{Review of Index-Free Formalism of \cite{spin} and \cite{blocks}\label{sec:spinreview}} The index-free formalism requires some familiarity with the embedding space or null cone approach to CFTs \cite{Weinberg:2010fx}. The idea of embedding space | one that hearkens back to Dirac \cite{1936dirac} | is now pervasive in the CFT literature. For this reason, we will omit reviewing this subject in great detail and instead direct readers who are unfamiliar with these ideas to the pedagogical review in \cite{slavanotes}. The discussion below is meant to bring the reader up to speed with the material needed for the $d=4$ calculation as quickly as possible, so the interested reader is also encouraged to consult the original literature vis-\'{a}-vis the index-free approach.

The essential idea of \cite{spin} is to encode the spin $\ell$ of a (symmetric) field by contracting it with vectors $z_{\mu_1} z_{\mu_2} \dots z_{\mu_\ell}$: \be
f(z) \equiv f_{\mu_1,\dots,\mu_\ell} z^{\mu_1} {}^{\dots} z^{\mu_\ell}.
\ee If the field has the added bonus of being traceless, then it may be recovered from $f(z)$ by restricting the polynomial to the region where $z^2=0$. This is seen from the fact that tracelessness implies $f(z)$ will be harmonic\footnote{This can be seen easily by computing $\frac{\partial^2 f(z)}{\partial z \cdot \partial z}$ and noting that the result will contain only contractions of the tensor $f$ with all its different indices.} and any polynomial may be written in the form of $h(z) + z^2 j(z)$, where $h(z)$ is harmonic, so $f(z)|_{z^2=0} = h(z)$. Another way of looking at this would be to note that a symmetric traceless tensor differs from a purely symmetric one by terms of $\mO(z^2)$. 

One can go further by lifting tensors to embedding space $f^{\mu_1,\mu_2,\dots, \mu_\ell}(x) \to F^{A_1, A_2, \dots, A_\ell}(P)$ where $F$ obeys the following essential conditions: \begin{enumerate} \item $F(\lambda P) = \lambda^{-\Delta} F(P)$: $F$ is degree $-\Delta$ in $P$, \item Defined where $P^2 = 0$, \item $P \cdot F = 0$: $F$ is transverse to $P_{A_i}$, \item $F$ is defined up to so-called ``pure gauge" terms, which are terms proportional to $P^{A_i}$. \end{enumerate} It is also implicit that $F$ must inherit the same symmetries of $f$ (that is, if $f$ is symmetric and traceless, $F$ will be as well). One can go back to physical space from embedding space by choosing the Poincar\'{e} section of the light cone $P^A = (1, x^2, x^\mu)$ and computing \be f_{\mu_1, \dots, \mu_{\ell}} = \pader{P^{A_1}}{x^{\mu_1}} \pader{P^{A_2}}{x^{\mu_2}} \dots \pader{P^{A_\ell}}{x^{\mu_\ell}} F_{A_1, \dots, A_{\ell}}. \ee Now, we perform the same trick as before in embedding space. We encode the (symmetric) tensor in terms of auxiliary vectors $Z^{A_1} \dots Z^{A_\ell}$: \be F(P;Z) \equiv F_{A_1, \dots, A_\ell} Z^{A_1} {}^{\dots} Z^{A_\ell}.\ee By analogy with the physical space picture, we may restrict traceless tensors to the region where $Z^2 = 0$. Moreover, since $F$ is defined up to pure gauge terms, we are also afforded the liberty to drop terms that are proportional to $Z \cdot P$. A last consistency relation is to note that since $F$ is of degree $-\Delta$ in $P$ and $Z\cdot P = 0$, there exists a shift symmetry for $F(P;Z) \to F(P;Z+\lambda P)$. Adhering to the above relations and sanity checks, one can construct the basic building blocks of two and three point functions of fields with arbitrary spin. 

Now, \cite{blocks} further extends the above index-free formalism to the conformal partial waves/blocks. For external scalar operators, the basic idea behind a conformal block differs very little in four dimensions compared to the two dimensional case. One can still insert the identity operator and organize according to irreducible representations of the conformal group i.e. determine the projection of a conformal family (primary and its descendants) to the four point function, which defines the block. Another way is to apply the OPE algebra \be \mO_1(x)\mO_2(y) \sim \sum_{\Delta, \ell} \lambda_{\Delta,\ell}  C(x-y,\partial_y) \mO_2(y), \ee  where $C(x-y,\partial_y)$ is determined completely by conformal invariance, twice to the four point function to obtain \begin{align} \langle \phi_1(x_1) \phi_2(x_2) \phi_3(x_3) \phi_4(x_4) \rangle &= \left(\frac{x_{24}^2}{x_{14}^2} \right)^{\frac{\Delta_1-\Delta_2}{2}}\left(\frac{x_{14}^2}{x_{13}^2} \right)^{\frac{\Delta_3-\Delta_4}{2}} \frac{1}{\left(x_{12}^2 \right)^{\frac{\Delta_1 + \Delta_2}{2}}\left(x_{34}^2 \right)^{\frac{\Delta_3 + \Delta_4}{2}}} \sum_{\Delta,\ell} \lambda_{\Delta,\ell}^{12}\lambda_{\Delta,\ell}^{34} G_{\Delta,\ell}(u,v) \nonumber \\
&\equiv \sum_{\Delta,\ell} \lambda_{\Delta,\ell}^{12}\lambda_{\Delta,\ell}^{34} W_{\Delta,\ell}(u,v), \end{align} where $u$ and $v$ are the conformally invariant cross ratios (see Appendix \ref{sec:cpwreview} for a lengthier review), given by \be u \equiv \frac{x_{12}^2 x_{34}^2}{x_{13}^2 x_{24}^2}, \,\,\,\,\,\,\,\,\,\,\,\,\,\,\,\,\,\,\,\, v \equiv \frac{x_{14}^2 x_{23}^2}{x_{13}^2 x_{24}^2}.\ee In four dimensions, the global conformal blocks $G_{\Delta,\ell}(u,v)$ (or equivalently, the partial waves $W_{\Delta,\ell}$) are given in terms of hypergeometric functions \cite{dolan}.  The idea of \cite{blocks} is to use this result to determine the blocks associated with external operators with spin. In effect, the spin structure of a correlation function is propagated by certain derivative operators acting on the scalar blocks. Naturally, this will subject our correlation functions to the same caveats that grant a closed form expression for the scalar blocks (namely, that the exchanged operators are symmetric and traceless). Since the four point function of scalars is obtained by gluing together three point functions of spin $(0,0,\ell)$, the task is to determine the ``left" and "right" differential operators that generate spinning operators $J_i$ out of scalar operators in the three point functions \begin{align} \langle J_1(P_1;Z_1) J_2(P_2;Z_2) \mO(P;Z) \rangle &= \mathcal{D}_{\textrm{left}} \langle \phi_1(P_1) \phi_2(P_2) \mO(P;Z) \rangle \\
\langle J_3(P_3;Z_3) J_4(P_4;Z_4) \mO(P;Z) \rangle &= \mathcal{D}_{\textrm{right}} \langle \phi_3(P_3) \phi_4(P_4) \mO(P;Z) \rangle, \end{align} such that the derivative operators propagate the index structure completely. The spin $(\ell_1, \ell_2, \ell_3)$ and $(0,0,\ell)$ three point functions can be determined in full generality. Combined with certain consistency conditions, the derivative operators can also be determined without ambiguity. They are: \begin{align} &D_{11} \equiv (P_1\cdot P_2)(Z_1\cdot \pader{}{P_2}) - (Z_1\cdot P_2)(P_1\cdot \pader{}{P_2}) - (Z_1\cdot Z_2)(P_1\cdot \pader{}{Z_2}) + (P_1\cdot Z_2)(Z_1\cdot\pader{}{Z_2}), \label{diff1} \\
&D_{12} \equiv (P_1\cdot P_2)(Z_1\cdot\pader{}{P_1}) - (Z_1\cdot P_2)(P_1\cdot\pader{}{P_1})+(Z_1\cdot P_2)(Z_1\cdot\pader{}{Z_1}),\\
&D_{21}\equiv (P_2\cdot P_1)(Z_2\cdot\pader{}{P_2}) - (Z_2\cdot P_1)(P_2\cdot\pader{}{P_2}) + (Z_2\cdot P_1)(Z_2\cdot \pader{}{Z_2}), \\
&D_{22}\equiv (P_2\cdot P_1)(Z_1\cdot \pader{}{P_1}) - (Z_2\cdot P_1)(P_2\cdot\pader{}{P_1}) - (Z_2\cdot Z_1)(P_2 \cdot \pader{}{Z_1}) + (P_2\cdot Z_1)(Z_2\cdot\pader{}{Z_1}). \label{diff2} \end{align} Along with one last trivial operator (in that does not affect the blocks) $H_{12}\equiv -2[(Z_1\cdot Z_2)(P_1\cdot P_2) - (Z_1\cdot P_2)(Z_2\cdot P_1)]$, these objects allow one to generate three point functions of spinning fields from a scalar-scalar-spin $\ell$ correlator. The notation denotes that the operator $D_{ij}$ raises the spin at point $i$ by one and lowers the scaling dimension at $j$ by one.

\subsubsection{Spin-1 Current\label{sec:spin1current}} We will now apply the techniques we reviewed in \S \ref{sec:spinreview} to the four point function consisting of a single spin-1 current and four other scalar operators, $\langle J_1(x_1) \mO_2(x_2) \mO_3(x_3) \mO_4(x_4) \rangle$. In the differential basis of Eqs. (\ref{diff1}) - (\ref{diff2}), this correlation function is a linear combination of \be D_{11} W_\mO^{10}, \,\,\,\,\,\,\,\,\,\,\,\,\,\,\,\,\,\,\,\,\,\,\,\, D_{12} W_\mO^{01}, \ee where $W_\mO^{ij}$ is the usual scalar conformal partial wave with $\Delta_1 \to \Delta_1 + i$ and $\Delta_2 \to \Delta_2 + j$. Note that the derivatives with respect to $Z_i$ vanish since the scalar partial waves only have dependence on the $P_i$. Therefore, acting on the scalar partial waves with the above derivative operators gives us \begin{align}D_{11}W_\mO^{10} &= (P_1\cdot P_2)\left[Z_1\cdot\pader{\chi^{10}}{P_2} G_\mO^{10} + \chi^{10} \left(\pader{G_\mO^{10}}{z}\lambda + \pader{G_\mO^{10}}{\zbar}\lbar \right) Z_1\cdot \pader{v}{P_2} + \chi^{10} \left( \pader{G_\mO^{10}}{z}\mu + \pader{G_\mO^{10}}{\zbar}\mbar \right) Z_1 \cdot \pader{u}{P_2} \right] \nonumber \\
&- (Z_1\cdot P_2)\left[Z_1\leftrightarrow P_1 \right], \label{d11} \\
D_{12} W^{01}_\mO &= (P_1\cdot P_2)\left[Z_1\cdot\pader{\chi^{01}}{P_1} G_\mO^{01} + \chi^{01} \left(\pader{G_\mO^{01}}{z}\lambda + \pader{G_\mO^{01}}{\zbar}\lbar \right) Z_1\cdot \pader{v}{P_1} + \chi^{01}\left( \pader{G_\mO^{01}}{z}\mu + \pader{G_\mO^{01}}{\zbar}\mbar \right) Z_1 \cdot \pader{u}{P_1} \right] \nonumber \\
&- (Z_1\cdot P_2)\left[Z_1\leftrightarrow P_1 \right], \label{d12}\end{align} where $\chi$ is the pre-factor of the partial waves ($W_\mO \equiv \chi G_\mO$), and $\mu$, $\mbar$, $\lambda$, and $\lbar$ are functions of $u$ and $v$ defined in Appendix \ref{sec:app1}. These functions arise because we have traded derivatives acting with respect to the $P_i$ in favor of the variables $z$ and $\zbar$, which are related to $u$ and $v$ via the relations: \begin{align} u &\equiv z\zbar, \\
v &\equiv (1-z)(1-\zbar).\end{align} In parity with the two dimensional case, the ET will concern how derivatives act on the scalar blocks, and we will devote the remainder of this section to this, although we have computed the full result in Appendix \ref{sec:divg1}. For notational simplicity, we further introduce  \be \phi^{ij,kl}_Y \equiv Y_k \cdot \pader{\chi^{ij}}{P_l},\,\,\,\,\,\,\,\,\,\,\, \zeta^{mn}_{Y,x} \equiv Y_m \cdot\pader{x}{P_n}. \ee With these definitions, let us focus on the action of $D_{11}$ first. Denoting $\tilde{(\cdot)}$ as the projection to physical space (meaning that we project all the $Z$'s and $P$'s to $z$'s and $x_i$'s), we have \begin{align} D_{11} W_\mO^{10} &= -\frac{1}{2}x_{12}^2\left[\tilde{\phi}^{10,12}_Z G_\mO^{10} + \chi^{10} \left(\pader{G_\mO^{10}}{z}\tilde{\lambda} + \pader{G_\mO^{10}}{\zbar}\tilde{\lbar} \right) \tilde{\zeta}^{12}_{Z,v} + \chi^{10} \left( \pader{G_\mO^{10}}{z}\tilde{\mu} + \pader{G_\mO^{10}}{\zbar}\tilde{\mbar} \right) \tilde{\zeta}^{12}_{Z,u} \right] \nonumber \\
&+ z_1 \cdot x_{12} \left(\tilde{\phi}^{10,12}_Z \leftrightarrow \tilde{\phi}^{10,12}_P, \tilde{\zeta}_Z \leftrightarrow \tilde{\zeta}_P \right). \label{eqn:d11} \end{align} We would like to calculate the divergence of the four point function. This is accomplished by first calculating $D_{11} W_{\mO}^{10}$ and $D_{12} W_{\mO}^{10}$, projecting to physical space, and then computing the action of the divergence $\mathcal{D}_c \equiv \pader{}{x}\cdot\pader{}{z}$ (for an explanation of why this works, see Appendix \ref{sec:appcons}). When we act with $\Dc$, there will be \textit{many} terms that are proliferated. It is therefore wise to systematically examine and categorize these terms in a qualitative way.  It turns out that all terms can be classified as follows: (a) double and single derivative terms of the form $\partial^2 G_\mO$ and $\partial G_\mO$ (the derivatives act with respect to $z$ or $\zbar$); (b) overall prefactors like $\tilde{\lambda}$ and $\tilde{\lbar}$ in Eq. (\ref{dd}) that are not relevant in determining if $\Dc \langle J \mO \mO \mO \rangle$ can be written in terms of scalar blocks; and (c) finite terms, which trivially become scalar functions multiplying scalar partial waves.

Two types of terms that will show up corresponding to category (a) are the double derivatives \be \tilde{\lambda}\dmpader{G_\mO}{x_1}{z}, \,\,\,\,\,\,\,\,\,\, \tilde{\lbar}\dmpader{G_\mO}{x_1}{\zbar}. \label{dd} \ee As we alluded to earlier, the $d=4$ the global conformal blocks are given in terms of hypergeometric functions, \begin{align}G_\mO(z,\zbar) &= \frac{(-)^l}{2^l} \frac{z\zbar}{z-\zbar}\left[k_{\Delta+l}(z)k_{\Delta-l-2}(\zbar) - (z\leftrightarrow \zbar) \right], \\
k_\beta(x)&\equiv x^{\beta/2} {_2 F_1}\left(\frac{\beta - \Delta_{12}}{2}, \frac{\beta + \Delta_{34}}{2}, \beta; x \right), \end{align} with $\Delta_{ij} \equiv \Delta_i - \Delta_j$. The problem of trying to figure out how these derivatives act on the scalar blocks can be translated into figuring out how they act on $k_\beta$. Consider the derivative acting with respect to $z$ first. Note that \begin{align} \pader{k_\beta}{z} &= \frac{z^{\beta /2} \left(\frac{2 \beta ^2 \, _2F_1\left(\frac{1}{2} \left(\beta -\Delta _{12}\right),\frac{1}{2} \left(\beta
   +\Delta _{34}\right);\beta ;z\right)}{z}+\left(\beta -\Delta _{12}\right) \left(\beta +\Delta _{34}\right) \,
   _2F_1\left(\frac{1}{2} \left(\beta -\Delta _{12}+2\right),\frac{1}{2} \left(\beta +\Delta _{34}+2\right);\beta
   +1;z\right)\right)}{4 \beta } \nonumber \\
   &= \frac{\beta}{2z}k_\beta(z) + \frac{(\beta-\Deot)(\beta+\Detf)}{4\beta \sqrt{z}} k_{\beta+1}^{\Delta_1 \to \Delta_1 - 1, \Delta_3 \to \Delta_3+1}(z), \end{align} and so we obtain what looks almost like a recursion relation \begin{align} \pader{G_{\Delta,l}}{z} &= -\frac{\zbar}{z(z-\zbar)}G_{\Delta,l} + \frac{(-)^l}{2^l} \frac{z\zbar}{z-\zbar} \bigg[\frac{(\Delta+l) k_{\Delta+l}(z)k_{\Delta-l-2}(\zbar) - (\Delta-l-2)k_{\Delta+l}(\zbar)k_{\Delta-l-2}(z)}{2z} \nonumber \\
   &+ M_{\Delta+l}(z) k_{\Delta-l-2}(\zbar)k_{\Delta+l+1}^{\Delta_1 \to \Delta_1 -1, \Delta_3 \to \Delta_3+1}(z) - M_{\Delta-l-2}(z) k_{\Delta+l}(\zbar)k_{\Delta-l-1}^{\Delta_1 \to \Delta_1 -1, \Delta_3 \to \Delta_3+1}(z)\bigg], \label{eqn:sder}\end{align} where we have defined $M_\beta(z) \equiv (\beta-\Deot)(\beta+\Detf)/(4\beta \sqrt{z}).$ The above result is a function of $z$ and $\zbar$ so we should write \be \dmpader{G_\mO}{x_1}{z} = \frac{\partial^2 G_\mO}{\partial z^2}\left(\tilde{\lambda}\pader{v}{x_1} + \tilde{\mu}\pader{u}{x_1} \right) + \dmpader{G_\mO}{\zbar}{z}\left(\tilde{\lbar}\pader{v}{x_1} + \tilde{\mbar}\pader{u}{x_1} \right), \ee so we have to consider the action of $\pader{}{z}$ and $\pader{}{\zbar}$ on Eq. (\ref{eqn:sder}). Let's first look at what happens when we apply $\pader{}{z}$ to $\pader{G}{z}$: \begin{align}
&\frac{\partial^2 G_{\Delta,l}}{\partial z^2} = \frac{\bar{z}}{4 z \left(z-\bar{z}\right)^2} \bigg\{8 G_{\Delta ,l}(z,\zbar) \nonumber \\
&+ \frac{(-)^l}{2^l} {\color{green} \bigg(}k_{a}(\bar{z}) {\color{red}\bigg[}4 z^2 \left(\bar{z}-z\right) M_{b}(z)
   M_{b+1}(z) k_{b+2}^{\Delta_1',\Delta_3'}(z)-2z\left(z-\bar{z}\right) (b+1) k_{b+1}(z) M_{b}(z)+ \nonumber \\
   &- k^{\Delta_1',\Delta_3'}_{b+1}(z) M_{b}(z)\left[z(b-1) + \zbar(b+3)\right] + b \left[\bar{z} (b+2)-z (b-2)\right]
   k_{b}(z){\color{red}\bigg]}\nonumber \\
   &+k_{b}(\bar{z}) {\color{blue}\bigg[} -4z^2 \left(\bar{z}-z\right) M_{a}(z) M_{a+1}(z) k_{a+2}^{\Delta_1',\Delta_3'}(z) +2z\left(z-\bar{z}\right) (a+1)
   k_{a+1}(z) M_{a}(z) \nonumber \\
   &+ k_{a+1}^{\Delta_1',\Delta_3'}(z) M_a(z)\left[z(a-1) + \zbar(a+3) \right]-a \left(\bar{z} (a+2)-z (a-2)\right) k_{a}(z){\color{blue} \bigg]}{\color{green} \bigg)}\bigg\}, \label{eqn:ddG}\end{align} where we have defined $a \equiv \Delta+l$ and $b \equiv \Delta-l-2$.
   
   The above result is one of three types of second-derivative terms that appears in $\Dc \langle J \mO \mO \mO \rangle$ (the other two should be the second derivatives in $\zbar$ and the mixed derivatives in $z$ and $\zbar$). Now, no approximations have been used so far, so the result we just obtained looks rather abstruse.  Evidently, there are explicit scalar blocks and there are other terms that do not resemble the scalar blocks at first glance. Again, in analogy with the two dimensional case, we can look at the large $\Delta$ limit (that is $\Delta \gg \Delta_1 \equiv \Delta_J$) of each term. This limit is interesting because we know that the twists $\tau = \Delta - l$ play the role of center-of-mass energies in the conformal partial waves. Therefore, when the center of mass energy is large for the 2 to 2 scattering in AdS, we expect the equivalence theorem to hold. Before we apply this limit to the above result, we first see that all the terms come in the with ``right" sign and can be grouped in the form \be k_a(z) k_b(\zbar) - k_a(\zbar) k_b(z).\ee However, there are two problems to deal with first: (1) we see that taking the large $\Delta$ limit is not enough to reproduce the scalar blocks (because $a$ and $b$ are different coefficients) and (2) there are shifted scalar blocks (shifted in the sense of $\Delta_1 \to \Delta_1-1$). The first problem can easily be dealt with by making a further assumption: $\Delta \gg l,1$ such that $a = \Delta + l \approx b = \Delta - l -2$. This simply translates to the fact that exchanged bulk fields do not carry large angular momentum, which would imply a large impact parameter and lower the overall energy carried by the gauge boson. For the ET theorem to hold, the energy carried by the bulk gauge boson must be large and the CFT reflects this property.
   
   

The second condition comes from the presence of shifted $k$-functions, when we interpreted $\pader{k_\beta}{z}$ as a recursion relation which relates the shifted $\beta \to \beta+1$ $k$-functions to $k_\beta$. The only way to do this was to assume that there was a unit dimension shift to either $\Delta_1$ \textit{and} $\Delta_3$ \textit{or} a unit dimension shift to $\Delta_2$ \textit{and} $\Delta_4$. However, because we choose $\Delta \gg 1$, these small shifts to the external operator dimensions can safely be absorbed into $\Delta$. The shifted terms then also combine in a simple way to produce scalar blocks in the large $\Delta$ limit. Another interpretation of this condition is to look at higher spin correlators. There, it turns out that the condition is really for $\Delta_{1,3} \gg s_{1,3}$ i.e. we do not consider an outlandish scenario where the twists of the external operators are dominated by their spins (this is simply because each derivative acting on a scalar block essentially becomes a factor of $\Delta$ in the large $\Delta$ limit and higher spin currents in correlation functions are written in the differential basis as more derivative operators acting on scalar partial waves).


Therefore, given the existence of the OPE where exchanged operators have twist $\tau = \Delta-\ell$ and are symmetric traceless, and the usual reasonable assumptions about a CFT$_4$, the following conditions are necessary and sufficient in order for $\Dc \langle J_1(x_1) \mO_2(x_2) \mO_3(x_3) \mO_4(x_4) \rangle$ (the scalar operators $\mO_i$ may all be different) to admit scalar modes consistent with the Goldstone equivalence theorem: \begin{enumerate} \item $\Delta \gg \Delta_{J_1} $, \item $\Delta \gg l$, $1$. \end{enumerate}
   
   
   
One might be concerned that so far that our analysis has only focused on the derivative terms that act with respect to $z$. There are certainly many more terms that appear in the divergence, but these terms do not refute the above result; the above conditions are necessary and sufficient. For example we have the following result for the $\zbar$ term: \begin{align}
 \dpader{G}{\zbar} &= \frac{1}{{4 \left(z-\bar{z}\right)^2 \bar{z}}} \bigg\{8 z G_{\Delta,l}(z,\bar{z})
+ \frac{(-)^l}{2^l} \bigg[\bar{z} k_b(z) \bigg(a^2 \left(\bar{z}-z\right) k_a(\zbar) +2\zbar {\color{blue} \big( }
   \left(\bar{z}-z\right) M_a(\zbar) {\color{red} (}(a+1) k_{a+1}(\zbar) \nonumber \\
   &+ 2 k_{a+2}^{\Delta_1',\Delta_3'} (\zbar) M_{a+1}(\zbar){\color{red})}+ k_{a+1}^{\Delta_1',\Delta_3'} (\zbar) M_a(\zbar)\left[\zbar(a+1) - z(a+5) \right]{\color{blue} \big)}\bigg) \nonumber \\
   &+ \bar{z} k_a(z) \bigg(-b^2 \left(\bar{z}-z\right) k_b(\zbar) -2\zbar {\color{blue} \big( }
   \left(\bar{z}-z\right) M_b(\zbar) {\color{red} (}(b+1) k_{b+1}(\zbar) \nonumber \\
   &+ 2 k_{b+2}^{\Delta_1',\Delta_3'} (\zbar) M_{b+1}(\zbar){\color{red})}+ k_{b+1}^{\Delta_1',\Delta_3'} (\zbar) M_b(\zbar)\left[\zbar(b+1) - z(b+5) \right]{\color{blue} \big)}\bigg) \nonumber \\
   &+ 4z\zbar\left( b k_b(\zbar) k_a(z) - a k_b(z) k_a(\zbar)\right) \bigg] \bigg\}.
   \end{align} The above result becomes a scalar block in the limits previously considered. There is also the mixed derivative term $\partial_z \partial_{\zbar} G_{\Delta,l}$ and it is straightforward to verify that it, too, adheres to this behavior (although of course there the result will contain a many more terms). We compute the full result of $\Dc\langle J_1(x_1) \mO_2(x_2) \mO_3(x_3) \mO(x_4) \rangle$ in Appendix \ref{sec:divg1} and verify that in the limit $\Delta \gg \Delta_J, l$ that the correlator reduces to the form \begin{align} \Dc \langle J_1(x_1) \mO_2(x_2) \mO_3(x_3) \mO(x_4) \rangle &\approx (f_1 + f_1')\sum_{\Delta,l} \lambda^{12}_{\Delta,l} \lambda^{34}_{\Delta,l} F_{1,\Delta} G_{\Delta,l} \nonumber \\
   &+ (f_2 + f_2') \sum_{\Delta,l} \lambda^{12}_{\Delta,l} \lambda^{34}_{\Delta,l}  F_{2,\Delta} G_{\Delta,l} + (f_3 + f_3') \sum_{\Delta,l} \lambda^{12}_{\Delta,l} \lambda^{34}_{\Delta,l}  F_{3,\Delta} G_{\Delta,l} \nonumber \\
&+ \textrm{ terms of order } \Delta \textrm{ and below},  \label{eqn:highDlim}\end{align} where $F_i$ are proportional to $\Delta^2$ (they are given explicitly in Appendix \ref{sec:divg1}) and also depend on the coordinates of the external operators through $z$ and $\zbar$. The idea then is that when one rescales the partial wave coefficients uniformly by $\lambda^{ij}_{\Delta,l} \to \Delta^{-1} \lambda^{ij}_{\Delta,l}$, then it is clear that the leading order terms will be \textit{scalar correlation} functions, modulo the prefactor functions. This is a subtle consequence of the conformal \textit{blocks} satisfying the ET.

To summarize, we found that \begin{itemize}
  \item the divergence of the four point function could be written in the differential basis as Eqs. (\ref{diff1}) - (\ref{diff2}).
  \item After some massaging of terms, the divergence could be expressed as \begin{align} \partial \cdot G_1 &= f_1 \partial^2 G^{10} + f_1' \partial^2 G^{01} + f_2 \partial \pbar G^{10} + f_2'\partial \pbar G^{01} \nonumber \\
&+ f_3 \pbar^2 G^{10} + f_3' \pbar^2 G^{01} + g_1 \partial G^{10} + g_1' \partial G^{01} + g_2 \pbar G^{10} + g_2' \pbar G^{01} + h G^{10} + h' G^{01},\end{align} where the derivatives are either with respect to $z$ $(\p)$ or $\zbar$ $(\pbar)$. The functions $f_i$, $g_i$, $h$ depend only on the dimensions and coordinates of the external operators. Their exact forms are given in Appendix \ref{sec:divg1}.
  \item In the large $\Delta$ limit, taking both $\Delta \gg l$, $\Delta_J$, we found that it reduces further to Eq. (\ref{eqn:highDlim}). The functions $F_{i,\Delta}$ are all $\sim \Delta^2$ to leading order and are also given in the Appendix.
  \item This implies that the blocks for the four point function become scalar blocks, assuming that the exchanged operators are symmetric and traceless. \end{itemize}

\section{Discussion\label{sec:disc}} We have examined the ET as a statement about propagating AdS massive gauge bosons and the corresponding CFT currents.

In the AdS bulk, we have defined an analogue to the $S$-matrix as a correlation function of creation and annihilation operators, and showed a relationship between such objects that involve $z$ ``polarized" gauge bosons (in the Poincar\'{e} patch) and their corresponding Goldstone boson when the magnitude of its momentum is large.  It was shown that these matrix elements naturally satisfy the AdS ET regardless of the external momentum scales when the conformal dimensions of the leading order exchanged particles are sufficiently larger than the dimension of the gauge boson.  This follows since arbitrarily increasing the exchanged scaling dimensions arbitrarily suppresses the interacting piece of the matrix element, which is compensated for by increasing the incoming momentum to the point that the equivalence theorem is satisfied anyway.  As a consequence, the divergence of conformal currents dual to the gauge boson in correlators for theories with ``heavy" exchange operators is approximately primary. Indeed, when the correlators are expressed as integrals of matrix elements, only the upper part of integration space contributes for the interacting pieces.

On the side of the CFT, we have shown that the conformal blocks for a correlator of a spin-1 current and three scalars satisfy the ET in the large twist limit. It would be a natural extension of this work to generalize the result to non symmetric tensors, making use of the ``shadow" formalism of \cite{davidsdshadow}, which offers an alternative method to treat spinning fields in full generality. 

It is also worthwhile to note that there are other interpretations of center of mass energies in a CFT. In flat space, the equivalence theorem is a statement of scattering amplitudes in a large momentum limit. For CFT correlation functions, the analog of momentum space is Mellin space (\cite{mellin1}, \cite{mellin2}, \cite{Penedones:2010ue}, \cite{Mack:2009mi}). It would be interesting to see if the ET can be obtained as a kinematic limit in Mellin space, following the derivation in flat space. 

In our treatment of CFT currents and conformal blocks, we did not mention higher spin broken currents in four dimensions. The reason why such an analysis is difficult is because the number of derivative terms needed to classify higher spin correlation functions grows  with the spin of the current. Therefore, computing higher spin correlation functions in terms of their conformal blocks becomes an exercise in computing many derivative expansions. In principle, one could use the index-free approach to understand the structure of higher spin broken currents (\cite{Maldacena:2011jn}, \cite{Maldacena:2012sf}). It would also be interesting, but difficult, to see show explicitly the presence of lower-spin terms at sub-leading order in $\Delta/\Delta_J$ in the conformal blocks for higher spin correlation functions.

\section{Acknowledgments} We thank Jared Kaplan for guidance and support throughout this project. NA also thanks Junpu Wang for useful discussions during the early stages of this work. NA and SC were supported in part by NSF grant PHY-1316665.

\appendix

\section{Spontaneously Broken Gauge Lagrangian}
Here, we justify the form of the Lagrangian in Eq. (\ref{lagrangian}).  In this section, overbars indicate a vector under a gauge group; the same symbol without an overbar is the magnitude of the corresponding vector. The `$\cdot$' symbol indicates a sum over all gauge indices; the `$\times$' symbol indicates a sum over only broken gauge indices; the `$*$' symbol indicates a sum over all gauge indices excluding the broken ones.

Consider an $SU(N)$ gauge theory coupled to a charged scalar, $\bar{\Phi}$, that acquires a vacuum expectation value, $\frac{\bar{v}}{\sqrt{2}}$: \begin{align}
\mathcal{L} \supset |(\partial - i g A\cdot T) \bar{\Phi}|^2 - V(|\Phi|^2). \label{lagrangian3}
\end{align} The Higgs mechanism breaks an $SU(N)$ symmetry down to an $SU(N-1)$ symmetry, resulting in $2N-1$ Goldstone bosons.  Since $\Phi$ originally had $2N$ degrees of freedom, only one Higgs-like degree of freedom remains.  $\Phi$ can thus be parameterized as \begin{equation}
\bar{\Phi} = \exp \left[i \frac{\pi}{v} \times T \right] (\bar{v}/\sqrt{2} + \bar{\phi}),
\end{equation} where the $\pi$'s are the Goldstone bosons.  Neither $\bar{v}$ nor $\bar{\phi}$ can be annihilated by the generators in the exponential, so we conclude $\bar{\phi} \propto \bar{v}$.

The kinetic term in Eq. (\ref{lagrangian3}) can then be written as \begin{align}
\left|(\partial - i g A \cdot T) e^{\left[i \frac{\pi}{v} \times T \right]} (\bar{v} + \bar{\phi}) \right|^2 =& \left| e^{\left[i \frac{\pi}{v} \times T \right]}  \partial \bar{\phi} - i g (A - m_A^{-1} \partial \pi) \times T e^{\left[i \frac{\pi}{v} \times T \right]} (\bar{v}/\sqrt{2} + \bar{\phi})    \right. \nonumber\\
&- \left. i g A * T e^{\left[i \frac{\pi}{v} \times T \right]}(\bar{v}/\sqrt{2} + \bar{\phi}) \right|^2 \\
=& \left| \partial \bar{\phi} - i g (A - m_A^{-1} D \pi) \times T  (\bar{v}/\sqrt{2} + \bar{\phi})   \right|^2 \\
=& \frac{1}{2} m_A^2 (A_{\text{broken}} - m_A^{-1} D \pi)^2 + \mathcal{L}_{int}(\frac{1}{2} (A_{\text{broken}} - m_A^{-1} D \pi)^2, \phi),
\end{align} where $m_A \equiv g v$.  The second line follows from the first since two generators from the broken part of $SU(N)$ commute into the preserved $SU(N-1)$ part, $[A\times T, e^{\left[i \frac{\pi}{v} \times T \right]}]^c =-m_A^{-1} g e^{\left[i \frac{\pi}{v} \times T \right]} A^a \pi^b f^{abc} T^c$ for structure constants $f^{abc}$, and $A *T \bar{v}=0$; the third line foliows since $T \times T = \mathbb{I}$.  The important upshot is that the Lagrangian for an $SU(N)$ gauge theory broken by the Higgs mechanism satisfies the form given in Eq. (\ref{lagrangian}).  This form holds for any spontaneous breaking mechanism and follows generally by simply demanding the Goldstone bosons transform as simple shifts, are derivatively coupled, and that gauge symmetry should still hold at the Lagrangian level.

\section{Review of the Schwinger-Dyson Equations} 

The following is an adaptation of the approach \cite{srednicki} takes to derive the Schwinger-Dyson equations. Consider a general path integral for some fields $\{ \phi_a\}$ of arbitrary spin in $d+1$ dimensions, \begin{align}
Z[\phi_a, J ] =& \int \mathcal{D} \phi_a e^{- i \left( S[\phi_a] - \int d^{d+1} x \,  J_a \cdot \phi_a \right) },
\end{align} where `$\cdot$' indicates a contraction of all indices between $J$ and $\phi$.  Now vary the fields in a manner commensurate with the path integral measure, $\phi_a \to \phi_a + \delta \phi_a$, and consider the resulting path integral, \begin{align}
Z[\phi_a + \delta \phi_a, J] =& \int \mathcal{D} \phi_a e^{- i \left( S[\phi_a] - \int d^{d+1} x  \,  J_a \cdot \phi_a \right) } e^{- i \left( \int d^{d+1} x \frac{\delta S[\phi_a]}{\delta \phi_a} \cdot \delta \phi_a  - \int d^{d+1} x \,  J_a \cdot \delta \phi_a \right) } \\
\approx &  \int \mathcal{D} \phi_a e^{- i \left( S[\phi_a] - \int d^{d+1} x  \,  J_a \cdot \phi_a \right) } \left[ 1- i  \int d^{d+1} x \left( \frac{ \delta S[\phi_a]}{\delta \phi_a}  - J_a \right)   \cdot \delta \phi_a \right] .
\end{align} Any transformation of fields that leaves the path integral measure invariant should leave the path integral itself invariant since fields are being integrated over all possible values anyway.  Since this was exactly our constraint on the transformation of the fields, we see \begin{align}
 Z[\phi_a + \delta \phi_a, J] &= Z[\phi_a, J] \\
\implies \int \mathcal{D} \phi_a e^{- i \left( S[\phi_a] - \int d^{d+1} x  \,  J_a \cdot \phi_a \right) } &\left[  \frac{ \delta S[\phi_a]}{\delta \phi_a}  -  J_a \right] = 0. \label{ds_equiv}
\end{align} Acting on both sides of Eq. (\ref{ds_equiv}) with $\prod_i \left( -i \frac{\delta}{\delta J_{a_i} (x_i)} \right)$ then setting $J_a=0$ yields  \begin{align}
 \left[\int \mathcal{D} \phi_a e^{- i S[\phi_a]  } \right] ^{-1} \int \mathcal{D} \phi_a e^{- i  S[\phi_a] } \frac{ \delta S[\phi_a]}{\delta \phi_a(x)} \prod_i \phi_{a_i}(x_i) =& \langle T  \frac{ \delta S[\phi_a]}{\delta \phi_a(x)}  \prod_i \phi_{a_i}(x_i) \rangle \nonumber \\
 &= -i  \sum_i \delta^{d+1} ( x - x_i). \label{schwinger_dyson}
\end{align}

By applying the differential operator associated with the classical equations of motion of a field to each field in a correlation function, the Schwinger-Dyson equations in Eq. (\ref{schwinger_dyson}) provide a tower of coupled differential equations that describe time ordered correlation functions sourced by contact terms (the delta functions).

Now consider the Lagrangian provided in Eq. (\ref{lagrangian2}).  The resulting Schwinger-Dyson equations for the broken gauge bosons and associated Goldstones are \begin{align}
[ \nabla_N \nabla^M - \xi^{-1} \nabla^M \nabla_N - ({\nabla^2}^M_N + m_A^2 \delta^M_N) ] \langle T A^{a N} \dots \rangle  =&  \langle T (A^{a M} - m_5^{-1} \partial^M \pi^a)\mathcal{L}'_{int} \dots \rangle \nonumber \\
& + \langle J^M \rangle + C_G \label{gauge_eom2}\\
(\nabla^2 + \xi m_A^2) \langle T  \pi^a \dots \rangle =  - m_A^{-1} \nabla_M &\langle T(A^{a M} - m_5^{-1} \partial^M \pi^a) \mathcal{L}_{int}' \dots \rangle \nonumber \\
&+ C_{GS}
\end{align}

where the $C$'s are contact terms, $\mathcal{L}'_{int}$ is the derivative of $\mathcal{L}_{int}$ with respect to its first argument, and $J^M$ is a conserved current to which the gauge fields couple, $J^{a M} =  \frac{\partial}{\partial {A^a}_M} \left[ \mathcal{L}_{G,int}[A^{a}] + \mathcal{L}_{GH}\right]$.  The `$\dots$' include other field operators.

Acting on both sides of Eq. (\ref{gauge_eom2}) with $m_A^{-1} \nabla_M$ annihilates the conserved current and yields Eq. (\ref{DS_upshot}):

\begin{equation}
m^{-1}_A \nabla_M[ \nabla_N \nabla^M - \xi^{-1} \nabla^M \nabla_N - ({\nabla^2}^M_N + m_A^2 \delta^M_N) ] \langle T A^{a N} \dots \rangle =  - [\nabla^2  + \xi m_5^2] \langle T  \pi^a \dots \rangle .
\end{equation} 

\section{Review of AdS Wave Functions}\label{wave_function_review}
It is well known that a (real) scalar field can be expanded in terms of eigenfunctions that satisfy the wave equation for the spacetime in which it dwells as \begin{align}
\phi(x) = \sum_i \left[ a_i f^\dagger_i(x) + h.c. \right],
\end{align} where \begin{align}
\left[ \nabla^2 + m_\phi^2 \right] f_i(x) = 0. \label{scalar_eqn}
\end{align} Here, $i$ is simply used as a schematic label for the eigenfunctions and can generally be discrete or continuous and represent many parameters.  In Poincar\'{e} patch coordinates in AdS the expansion becomes what is seen in Eq. (\ref{scalar_expansion}) with wave functions given by  \begin{align}
f(\vec{p}, m) =& N_\phi z^{\frac{d}{2}} J_{\Delta_\phi -\frac{d}{2}} (m z) e^{-i p_m \cdot x}
\end{align} where $N_\phi$ is a normalization factor.  Defining the inner product on function space over constant-time slices of $AdS$ as \begin{align} \label{scalar_product}
\langle \Psi_1 , \Psi_2 \rangle = i \int d^d x \, \sqrt{ g} \,  g^{0 0 } \left [ \Psi_1^\dagger \partial_0 \Psi_2 - \partial_0 \Psi^\dagger_2 \Psi_1 \right]
\end{align} and demanding $\langle f_m ,  f_n \rangle = \delta_{mn}$ yields the normalization $N_\phi= \frac{\sqrt{m}}{\sqrt{2 p_{m 0}}}$, resulting in wave functions given by Eq. (\ref{scalar_ads_functions}).


Gauge fields can be expanded in a similar manner, but solving for their corresponding wave functions is slightly more involved. Since $A\in \mathcal{H} \otimes \mathcal{H}^* \otimes \mathbb{R}^{d+1} \otimes \left[ C^2 \right]^{d+1}$ is Hermitian, we can always expand the gauge field in terms of eigenfunctions as: \begin{align}
A^a_M = \sum_{s,i} \left[ a^a_{s,i} h^\dagger_{s,i M} (x) + h.c. \right], \label{gauge_expansion}
\end{align} where $a_{s,i} \in \mathcal{H} \otimes \mathcal{H}^* $ and $h_{s,i M} (x)\in  \mathbb{C}^{d+1} \otimes \left[ C^2 \right]^{d+1}$.  The index `$s$' labels our basis in $\mathbb{C}^{d+1}$, thereby taking a value in a discrete, finite set.  Demanding that the fields satisfy the Heisenberg equations of motion, we must choose $h_{s,i M}$ that have the same Casimir weight as the free fields.  They must thus satisfy the free classical equations of motion for a massive gauge field: \begin{align}
\left[ \nabla_N \nabla^M - \xi^{-1} \nabla^M \nabla_N - \left({\nabla^2} + m_A^2  \right) \delta^M_N \right] h^N_{s,i}(x) = 0 \label{vec_eqn}.
\end{align} Taking the covariant divergence of Eq. (\ref{vec_eqn}) and defining $\tilde{h}_{s,i}(x) \equiv (\xi m_A)^{-1} \nabla_M h^M_{s,i}(x)$ yields \begin{align}
\left[ \nabla^2 + \xi m_A^2 \right] \tilde{h}_{s,i}(x) =0, \label{divergence_eqn}
\end{align} so the divergence of the vector wave function obeys the scalar wave equation.  The vector wave function with the divergence degree of freedom projected out, $\bar{h}_{s,i M}(x) \equiv \left( \delta^N_M - \partial_M \nabla^{-2} \nabla^N \right) h_{s,i M}(x)$, then satisfies \begin{align}
\left[ \nabla_N \nabla^M - \left( {\nabla^2}^M_N + m_A^2  \delta^M_N \right) \right] \bar{h}^N_{s,i}(x) = 0. \label{divergenceless_eqn}
\end{align} The full wave function is then constructed by solving Eqs. (\ref{divergence_eqn}) and (\ref{divergenceless_eqn}) and writing $h_{s,i}^M (x) = \partial^M \nabla^{-2} \tilde{h}_{s,i}(x) + \bar{h}_{s,i}^M (x)$.

Since the set $\{h_s\}$ is linearly independent and each $h_s$ can be decomposed into a linear combination of divergenceless degrees of freedom and a scalar divergence, we can select the set such that $s=\xi$ contains only the scalar degree of freedom and all others are divergenceless.  Additionally, we can select the set such that among the divergenceless wave functions only $h_z$ has a nonvanishing $z$-component. Under this prescription, the operator in Eq. (\ref{divergenceless_eqn}) can be diagonalized, leading to the expansion of the gauge fields to become what is seen in Eq. (\ref{vector_expansion}) with wave functions given by Eqs. (\ref{vector_ads_functions}) and (\ref{z_ads_function}): \begin{align}
h_{s, M} (\vec{p}, m) \underset{s \ne z, \xi}=& \left\{ \begin{matrix}
0, & M=z\\
N_A \epsilon_{s, \mu}(\vec{p}) z^{\frac{d}{2} -1 } J_{\Delta - \frac{d}{2}} (m z) e^{- i p_m \cdot x}, & M=\mu
\end{matrix}
\right. , \\
h_{z, M} (\vec{p}, m) &= \left\{ \begin{matrix}
N_A z^{\frac{d}{2}} J_{\Delta - \frac{d}{2}} (m z) e^{-i p_m \cdot x}, & M=z \\
- i N_A \frac{p_{m \mu}}{m^2}     \left[ m z^{\frac{d}{2}} J_{\Delta - \frac{d}{2} + 1 } (m z) - [\Delta - (d - 1) ] z^{\frac{d}{2} - 1} J_{\Delta - \frac{d}{2}} (m z) \right] e^{ - i p_m \cdot x}, & M=\mu
\end{matrix}
\right. .
\end{align}

Defining the inner product on vector function space as

\begin{align} \label{vector_product}
\langle \Psi_1 , \Psi_2 \rangle = i \int d^d x \, \sqrt{g} \, g^{0 0} \bigg\{& g^{M N}  \left[ \left(\Psi^\dagger_{1 M}  \nabla_0 \Psi_{2 N} - \nabla_0 \Psi^\dagger_{1 M} \Psi_{2 N}   \right)     -   \left( \Psi^\dagger_{1 M} \nabla_N \Psi_{2 0} - \nabla_M \Psi^\dagger_{1 0} \Psi_{2 N} \right) \right] \nonumber \\
&       + \xi^{-1}  \left( \Psi_{1 0}^\dagger \nabla^M \Psi_{2 M} - \nabla^M \Psi^\dagger_{1M} \Psi_{2 0} \right) \bigg\}
\end{align} and demanding $\langle h_{s, i}, h_{s', j} \rangle = \delta_{s s'} \delta_{ij}$ yields the normalization $N_A = \frac{\sqrt{m}}{\sqrt{2 p_{m 0}}}$, resulting in vector wave functions given by Eqs. (\ref{vector_ads_functions}) and (\ref{z_ads_function}).

\section{Review of the LSZ Formula \label{lsz_section}} 

Obtaining the ET in AdS required us to consider matrix elements and exploit the upshot of the Schwinger-Dyson equations within an LSZ integral.  To determine the LSZ-like integral on a curved spacetime, consider the following simple matrix element, \begin{align} \langle T a_{s, \vec{p},m}(+\infty) a^\dagger_{s, \vec{p},m}(-\infty) \rangle. \label{scattering_example}
\end{align} To find a functional form of Eq. (\ref{scattering_example}), we write \begin{align}
&a_{s, \vec{p},m} (+\infty) = a_{s, \vec{p},m}( -\infty) + \left[ \langle h_{s, \vec{p}, m}(+\infty) , A(+\infty) \rangle - \langle h_{s, \vec{p}, m}(-\infty) , A(-\infty) \rangle  \right] \nonumber \\
&= a_{s, \vec{p}, m} (- \infty) + i \int_{-\infty}^{+\infty} dt \int d^d x \, \sqrt{g} \,  h_{s,\vec{p},m, M}^\dagger  \left[  \left( {\nabla^2}^M_N + m_A^2 \delta^M_N \right) - \nabla_N \nabla^M + \xi^{-1} \nabla^M \nabla_N \right]  A^N, \label{LSZ2}
\end{align} where the integral expression in the second line follows from the first by using the definition of the vector function space inner product in Eq. (\ref{vector_product}) and noting that the differential operator appearing in the second line annihilates $h_{s, \vec{p}, m}$.

We find a similar expression for $a^\dagger_{s, \vec{p}, m} ( -\infty)$: \begin{align}
a^\dagger_{s, \vec{p}, m} ( -\infty ) = a^\dagger_{s, \vec{p}, m} ( +\infty) - i \int_{-\infty}^{+\infty} dt  \int d^d x \, \sqrt{g} \,  h_{s,\vec{p},m M}  \left[  \left( {\nabla^2}^M_N + m_A^2 \delta^M_N \right) - \nabla_N \nabla^M + \xi^{-1} \nabla^M \nabla_N \right]  A^N. \label{LSZ3}
\end{align} Inserting Eqs. (\ref{LSZ2}) and (\ref{LSZ3}) back into Eq. (\ref{scattering_example}) yields \begin{align}
\langle T  a_{s, \vec{p},m}(+\infty) a^\dagger_{s, \vec{p},m}(-\infty) \rangle =& (i)^2 \int dx \, dx' \, \sqrt{g(x) g(x')} \,  h_{s,\vec{p},m M}(x) h^\dagger_{s,\vec{p},m M'}(x')  \nonumber \\
&\left[  \left( {\nabla^2}^M_N + m_A^2 \delta^M_N \right) - \nabla_N \nabla^M + \xi^{-1} \nabla^M \nabla_N \right]^2 \langle T A(x) A(x') \rangle. \label{LSZ_formula1} 
\end{align} 

Multiplying by the appropriate state normalization $N=\sqrt{2 p_{m 0} m}$ allows us to write a variant of Eq. (\ref{LSZ_formula1}) that respects AdS isometries, \begin{align}
\langle s, \vec{p}, m | s, \vec{p}, m \rangle =& (i)^2 \int dx \, dx' \, \sqrt{g(x) g(x')} \,  2 p_{m 0} m h_{s,\vec{p},m M}(x) h^\dagger_{s,\vec{p},m M'}(x')  \nonumber \\
&\left[  \left( {\nabla^2}^M_N + m_A^2 \delta^M_N \right) - \nabla_N \nabla^M + \xi^{-1} \nabla^M \nabla_N \right]^2 \langle T A(x) A(x') \rangle. \label{LSZ_formula2}
\end{align} This is the LSZ reduction formula that relates correlation functions of creation and annihilation operators to correlation functions of fields.  The differential operator acting on the correlation function in the integrand of Eq. (\ref{LSZ_formula2}) generates contact terms that correspond to disconnected diagrams according to the Schwinger-Dyson equations.

The above steps can be repeated for any number of $a$'s to arrive at, for instance, Eq. (\ref{ads_lsz_exchange}).

\section{Review of Conformal Blocks and Partial Waves\label{sec:cpwreview}} As we have emphasized, there are two ways of looking at the (global) conformal blocks. One is purely algebraic | we simply apply the OPE algebra twice and use the orthogonality of the two point function. The other approach is to insert the identity operator and then organize the contribution to the four point function in terms of the representations under the conformal group, namely the spins and scaling dimensions of the exchanged operators. In both cases, the end result is the same |  we can break up the structure of the correlation function into a sum over all conformal families of the theory. The dynamics are fully encoded in the coefficients of these operators while the blocks themselves only depend on the conformally invariant cross-ratios $u$ and $v$. In other words, for a primary operator appearing in our theory, the conformal blocks tell us how much that primary and its descendants contribute to the four-point function.

 Consider the four-point function of scalar fields $\phi_i \equiv \phi(x_i)$ with scaling dimensions $\Delta_i$. We can decompose it into an overall conformally invariant structure multiplied by some generic function of the conformal cross ratios $G(u,v)$ as follows: \be \langle \phi_1\phi_2\phi_3\phi_4 \rangle = \left(\frac{x_{24}^2}{x_{14}^2} \right)^{\frac{\Delta_1-\Delta_2}{2}}\left(\frac{x_{14}^2}{x_{13}^2} \right)^{\frac{\Delta_3-\Delta_4}{2}} \frac{1}{\left(x_{12}^2 \right)^{\frac{\Delta_1 + \Delta_2}{2}}\left(x_{34}^2 \right)^{\frac{\Delta_3 + \Delta_4}{2}}}G(u,v), \ee with $u\equiv \frac{x_{12}^2 x_{34}^2}{x_{13}^2 x_{24}^2}$ and $v \equiv \frac{x_{14}^2 x_{23}^2}{x_{13}^2 x_{24}^2}$. We can then express $G(u,v)$ as an expansion of functions (the conformal blocks) but noting that applying the OPE on the left hand side twice gives us \begin{align} \langle \phi_1\phi_2\phi_3\phi_4 \rangle &= \langle \sum_{\tau,\ell} \lambda^{12}_{\tau,\ell} C(x_1-x_2;\partial_2) \mO_2 \sum_{\tau',\ell'} \lambda^{34}_{\tau',\ell'} C(x_3-x_4;\partial_4) \mO_4 \rangle \nonumber \\
&=\sum_{\tau,\ell}\lambda^{12}_{\tau,\ell} \lambda^{34}_{\tau,\ell}  C(x_1-x_2;\partial_2)C(x_3-x_4;\partial_4) \langle \mO_2 \mO_4 \rangle \nonumber \\
&\equiv \left(\frac{x_{24}^2}{x_{14}^2} \right)^{\frac{\Delta_1-\Delta_2}{2}}\left(\frac{x_{14}^2}{x_{13}^2} \right)^{\frac{\Delta_3-\Delta_4}{2}} \frac{1}{\left(x_{12}^2 \right)^{\frac{\Delta_1 + \Delta_2}{2}}\left(x_{34}^2 \right)^{\frac{\Delta_3 + \Delta_4}{2}}} \sum_{\tau,\ell} \lambda_{\tau,\ell}^{12}\lambda_{\tau,\ell}^{34} G_{\tau,\ell}(u,v) \label{eqn:4ptcpw},\end{align} where we have used the fact that the two point function demands $\delta_{\tau,\tau'}\delta_{\ell,\ell'}$ due to orthogonality. The conformal blocks are denoted by $G_{\tau,\ell}(u,v)$ and the partial waves are defined as the blocks times additional coordinate-dependent prefactors: \be W_{\Delta, \ell} \equiv \left(\frac{x_{24}^2}{x_{14}^2} \right)^{\frac{\Delta_1-\Delta_2}{2}}\left(\frac{x_{14}^2}{x_{13}^2} \right)^{\frac{\Delta_3-\Delta_4}{2}} \frac{1}{\left(x_{12}^2 \right)^{\frac{\Delta_1 + \Delta_2}{2}}\left(x_{34}^2 \right)^{\frac{\Delta_3 + \Delta_4}{2}}}  G_{\tau,\ell}(u,v). \ee Note then that one can obtain the function $G(u,v)$ from the conformal blocks via \be G(u,v) = \sum_{\tau,\ell} \lambda^{12}_{\tau,\ell} \lambda^{34}_{\tau,\ell} G_{\tau,\ell}(u,v), \ee for the case of scalar fields. We sometimes abbreviate $G_{\tau,\ell}$ as $G_\mO(u,v)$ and the sum therefore runs over $\mO$.

\section{Details and Definitions \label{sec:app1}} \subsection{Partial Wave Definitions} To account for possible dimension shifts, we define the scalar partial waves as a scalar part ($\chi$) times the scalar block: \be W^{ij}_\mO \equiv \chi^{ij}G_\mO^{ij},\ee with \be \chi^{ij} \equiv \frac{1}{P_{12}^{\frac{1}{2}(\Delta_1 + i + \Delta_2 + j)} P_{34}^{\frac{1}{2}(\Delta_3+\Delta_4)}} \left(\frac{P_{24}}{P_{14}} \right)^{\frac{1}{2}(\Delta_1 + i - \Delta_2 - j)} \left(\frac{P_{14}}{P_{13}} \right)^{\frac{1}{2}\Delta_{34}}, \ee where $\Delta_{ij} \equiv \Delta_i - \Delta_j$. To obtain the partial wave in physical space, one may use the relation $P_{ij} = x_{ij}^2$.

\subsection{Scalar Functions} The scalar functions result from writing derivative operators acting on $z$ and $\zbar$ instead of $P_1$ and $P_2$. The point is that although $u$ and $v$ are the ``physical" variables of the scalar blocks, their exact forms are hypergeometric functions in $z$ and $\zbar$, which carry the implicit dependence on $u$ and $v$. We can write \be \pader{G_\mO}{P_i^A} = \pader{G}{z}\left(\pader{v}{P_i^A}\pader{z}{v} + \pader{u}{P_i^A}\pader{z}{u} \right) + \pader{G}{\zbar}\left(\pader{v}{P_i^A}\pader{\zbar}{v} + \pader{u}{P_i^A}\pader{\zbar}{u} \right), \ee where the partial derivatives acting on $\zbar$ are somewhat complicated since the condition that $u=z\zbar$ and $v=(1-z)(1-\zbar)$ lets us solve explicitly for $z$ and $\zbar$ in terms of $u$ and $v$ (there are two solutions):
\begin{align} &\left\{z\to \frac{1}{2} \left(-\sqrt{(u-v+1)^2-4 u}+u-v+1\right),\zbar \to \frac{1}{2} \left(\sqrt{(u-v+1)^2-4
   u}+u-v+1\right)\right\}, \\
   &\left\{z\to \frac{1}{2} \left(\sqrt{(u-v+1)^2-4 u}+u-v+1\right),\zbar \to \frac{1}{2}
   \left(-\sqrt{(u-v+1)^2-4 u}+u-v+1\right)\right\}. \label{eqn:zsoln}\end{align} If we interpret $z$ and $\zbar$ as coordinates then in the limit that $u,v \ll 1$, we find that in the first solution $z \to 0$ and $\zbar \to 1$ while for the second solution, $z \to 1$ and $\zbar \to 0$. Typically, in a four point function we can always do a rescaling of the external coordinates so we have $\langle \phi(0)\phi(z)\phi(1)\phi(\infty) \rangle$ which then implies $u \sim x_{12}^2 \sim z\zbar$. This means that which solution we pick is not of great importance since the correlation function only depends on the absolute distance between operators. Picking the second solution, we get \begin{align} \mu(u,v) &\equiv \pader{z}{u} = \frac{1}{2} \left(\frac{u-v-1}{\sqrt{(u-v+1)^2-4 u}}+1\right), \label{eqn:mu1} \\
   \bar{\mu}(u,v) &\equiv \pader{\zbar}{u} = \frac{1}{2} \left(\frac{-u+v+1}{\sqrt{(u-v+1)^2-4 u}}+1\right), \\
   \lambda(u,v) &\equiv \pader{z}{v} = \frac{1}{2} \left(\frac{-u+v-1}{\sqrt{(u-v+1)^2-4 u}}-1\right), \\
   \bar{\lambda}(u,v) &\equiv \pader{\zbar}{v} = \frac{1}{2} \left(\frac{u-v+1}{\sqrt{(u-v+1)^2-4 u}}-1\right). \label{eqn:lbar} \end{align}

\section{Applying Conservation\label{sec:appcons}} Recall that in position space, correlation functions involving currents are encoded into the $z$'s (not to be confused with $z$ and $\zbar$ that appear in the scalar blocks). Namely, for a correlator $f^{\mu_1 \mu_2 \dots \mu_n}$, we have \be \tilde{f}(x;z) \equiv f^{\mu_1 \mu_2 \dots \mu_n}(x) z_{1, \mu_1} z_{2, \mu_2} \dots z_{n, \mu_n}. \ee By lifting this to embedding space, we can recover the correlator in terms of the $Z$'s and $P$'s we have been using this whole time. However, it's easy to project back onto position space via the relations \begin{align} &Z_1 \cdot Z_2 \to z_1\cdot z_2, \,\,\,\,\,\,\,\,\,\,\,\,\,\,\,\, P_1\cdot P_2 \to -\frac{1}{2} x_{12}^2 \label{proj1} \\
&P_1\cdot Z_2 \to z_2\cdot x_{12}, \,\,\,\,\,\,\,\,\,\,\,\,\,\,\,\, P_2\cdot Z_1 \to -z_1 \cdot x_{12}. \label{proj2} \end{align} With this in mind, consider a two-point correlation function given by \be \tilde{f}(x;z) = {f}^{\mu\nu}(x) z_{1,\mu} z_{2, \nu}. \ee So we see that the $\partial_\mu$ operator is implemented easily through the \be \Dc \equiv \pader{}{x}\cdot\pader{}{z} \ee operator. One question might be why can't we do it in embedding space? Well, the answer is in principle one could do that noting that we can transform the above operator into partial derivatives acting on $P$'s and $Z$'s. The unfortunate price to pay would be to keep track of tensors like $\pader{Z^A}{z}.$ It is thus simpler to project onto physical space via the relations Eqs. (\ref{proj1}) - (\ref{proj2}) and then implement $\partial_x \cdot \partial_z$. It is easy to see that this argument generalizes quite readily to any $n$-point function as well, since evaluating the divergence at $x_i$, $z_i$ will always amount to computing $\partial_\mu f^{\mu \dots}$, modulo pre-factors of $z$ that clearly cannot influence the conservation condition.

\section{Full Result of Spin-1 Divergence\label{sec:divg1}} Let $f_1$ be coefficient function of $\partial^2 G$, $f_2$ for $\partial \bar{\partial} G$, and $f_3$ for $\pbar^2 G$. And for single derivatives, $g_1$ for $\partial G$ and $g_2$ for $\pbar G$. Lastly, we denote $h$ for the ``finite" term. The divergence of the four point function is then written as \begin{align} \partial \cdot G_1 &= f_1 \partial^2 G^{10} + f_1' \partial^2 G^{01} + f_2 \partial \pbar G^{10} + f_2'\partial \pbar G^{01} \nonumber \\
&+ f_3 \pbar^2 G^{10} + f_3' \pbar^2 G^{01} + g_1 \partial G^{10} + g_1' \partial G^{01} + g_2 \pbar G^{10} + g_2' \pbar G^{01} + h G^{10} + h' G^{01},\end{align} where $G_1$ is the single current four point function and the $G^{ij}$'s are scalar blocks with dimension shifts corresponding to $\Delta_1 \to \Delta_1 + i$ and $\Delta_2 \to \Delta_2 + j$. The crux of our analysis is that at large $\Delta$, $k$ derivatives acting on the scalar blocks become $\Delta^k G$ but this involves new scalar functions. At large $\Delta$, we found that \begin{align} \partial \cdot G_1 &\approx (f_1 + f_1')\sum_{\Delta,l} \lambda^{12}_{\Delta,l} \lambda^{34}_{\Delta,l} F_{1,\Delta} G_{\Delta,l} + (f_2 + f_2') \sum_{\Delta,l} \lambda^{12}_{\Delta,l} \lambda^{34}_{\Delta,l}  F_{2,\Delta} G_{\Delta,l} + (f_3 + f_3') \sum_{\Delta,l} \lambda^{12}_{\Delta,l} \lambda^{34}_{\Delta,l}  F_{3,\Delta} G_{\Delta,l} \nonumber \\
&+ \textrm{ terms of order } \Delta \textrm{ and below}, \end{align} where the functions $F_{i,
\Delta}(z,\zbar)$ came about from computing the double derivatives and taking the large $\Delta$ limit (for example, $F_{1,\Delta}$ would be the all the factors of $z$ and $\zbar$ in front of the scalar blocks in Eq. (\ref{eqn:ddG}) in the large $\Delta$ limit). To leading order in $\Delta$, the functions $F_{i,\Delta}(z,\zbar)$ are all proportional to $\Delta^2$. If one uniformly rescales the partial wave coefficients such that $\lambda^{ij}_{\Delta,l} \to \Delta^{-1} \lambda^{ij}_{\Delta,l}$, then it is evident that one obtains scalar correlation functions to leading order.  Here, we explicitly write down the $f_i$, $g_i$, and $h$ functions and $F_1$, $F_2$, $F_3$. \begin{align} f_1 &= c_1 \chi^{10} \bigg\{-\frac{1}{2} x_{12}^2 \left[2v\left(\frac{x_{13,\mu}}{x_{23}^2} - \frac{x_{14,\mu}}{x_{24}^2} \right)\left(\lambda^2 \partial^\mu v + \lambda \mu \partial^\mu u \right) - {\color{black} \frac{2u x_{14,\mu}}{x_{24}^2} \left(\mu\lambda \partial^\mu v + \mu^2 \partial^\mu u \right) } \right] \nonumber \\
&- \frac{2v x_{14,\mu}}{x_{24}^2}\left( \mu^2 \partial^\mu u + \mu \lambda \partial^\mu v \right) \bigg\},  \\
f_1' &= c_2\chi^{01}\bigg\{-\frac{1}{2}x_{12}^2\left[2v\left(\frac{x_{14,\mu}}{x_{14}^2} - \frac{x_{13,\mu}}{x_{13}^2} \right)\left(\lambda^2 \partial^\mu v + \mu\lambda \partial^\mu u \right) + 2u\left(\frac{x_{12,\mu}}{x_{12}^2} - \frac{x_{13,\mu}}{x_{13}^2} \right)\left(\mu \lambda \partial^\mu v + \mu^2 \partial^\mu u \right) \right] \bigg\}, \\
f_2 &= c_1\chi^{10} \bigg\{ -\frac{1}{2} x_{12}^2\left[2v\left(\frac{x_{13,\mu}}{x_{23}^2} - \frac{x_{14,\mu}}{x_{24}^2} \right)\left(2\lambda \lbar \partial^\mu v + \lambda \mbar \partial^\mu u + \lbar \mu \partial^\mu u \right) - {\color{black} \frac{2u x_{14,\mu}}{x_{24}^2} \left(\mu \lbar \partial^\mu v + \mbar \lambda \partial^\mu v + 2\mbar \mu \partial^\mu u \right) } \right] \nonumber \\
&- \frac{2v x_{14,\mu}}{x_{24}^2}\left(2\mu \mbar \partial^\mu u +\mu \lbar \partial^\mu v  + \mbar \lambda \partial^\mu v \right) \bigg\}, \\
f_2' &= c_2\chi^{01}\bigg\{-\frac{1}{2}x_{12}^2 \bigg[2v\left(\frac{x_{14,\mu}}{x_{14}^2} - \frac{x_{13,\mu}}{x_{13}^2}  \right)\left(2\lambda \lbar \partial^\mu v + \mbar \lambda \partial^\mu u + \mu \lbar \partial^\mu u\right) \nonumber \\
&+ 2u\left(\frac{x_{12,\mu}}{x_{12}^2} - \frac{x_{13,\mu}}{x_{13}^2}  \right)\left(\mu \lbar \partial^\mu v + \mbar \lambda \partial^\mu v + 2\mbar \mu \partial^\mu u \right) \bigg] \bigg\}, \\
f_3&= f_1(\lambda \leftrightarrow \lbar, \mu \leftrightarrow \mbar), \\
f_3'&= f_1'(\lambda \leftrightarrow \lbar, \mu \leftrightarrow \mbar),\end{align}


\begin{align}
g_1 &= c_1\chi^{10}\bigg\{- \frac{1}{2} x_{12}^2{\color{red}\bigg[} \frac{2\gamma}{x_{24}^2} x_{14,\mu}\left(\lambda \partial^\mu v + \mu \partial^\mu u\right) + 2v\partial_\mu \lambda \left(\frac{x_{13}^\mu}{x_{23}^2} - \frac{x_{14}^\mu}{x_{24}^2} \right) \nonumber \\
&+ 4\left(\alpha \frac{x_{12}^\mu}{x_{12}^2} + (\gamma-k)\frac{x_{14}^\mu}{x_{14}^2} + k\frac{x_{13}^\mu}{x_{13}^2} \right)\left[\lambda v\left(\frac{x_{13,\mu}}{x_{23}^2} - \frac{x_{14,\mu}}{x_{24}^2} \right) -\mu u \frac{x_{14,\mu}}{x_{14}^2} \right] \nonumber \\
&+2\lambda\left[\partial^\mu v \left(\frac{x_{13,\mu}}{x_{23}^2} - \frac{x_{14,\mu}}{x_{24}^2} \right) + vd \left(\frac{1}{x_{23}^2} - \frac{1}{x_{24}^2}\right) \right] - 2u\frac{x_{14}\cdot\partial \mu}{x_{24}^2} - 2\frac{\mu d v}{x_{24}^2}{\color{red}\bigg]} \nonumber \\
&+ 2vx_{12}^\mu \left[ \left(\frac{x_{13,\mu}}{x_{23}^2} - \frac{x_{14,\mu}}{x_{24}^2} \right) \lambda - \frac{2 x_{14,\mu}}{x_{24}^2} \mu\right] \nonumber \\
&+ 2\left(\alpha + (\gamma-k)\frac{x_{12}\cdot x_{14}}{x_{14}^2} + k\frac{x_{12}\cdot x_{13}}{x_{13}^2} \right)\left[\lambda v\left(\frac{x_{13}^2}{x_{23}^2} - \frac{x_{14}^2}{x_{24}^2} \right) -\frac{x_{14}^2}{x_{24}^2}u \mu \right] \bigg\}, \\
g_1'&= c_2 \chi^{01} \bigg\{ -\frac{1}{2} x_{12}^2 {\color{red}\bigg[} -2\left(\alpha\frac{x_{12,\mu}}{x_{12}^2} + \frac{x_{14,\mu}}{x_{14}^2}(\gamma'-k) + k\frac{x_{13,\mu}}{x_{13}^2} \right){\color{blue}\bigg(} \lambda \partial^\mu v + \mu \partial^\mu u  -2u\mu\left(\frac{x_{12,\mu}}{x_{12}^2} - \frac{x_{13,\mu}}{x_{13}^2} \right) \nonumber \\
&-2v\lambda\left(\frac{x_{14,\mu}}{x_{14}^2} - \frac{x_{13,\mu}}{x_{13}^2} \right) {\color{blue}\bigg)}  {\color{red}\bigg]} + 2\lambda \left[\partial^\mu v \left(\frac{x_{14,\mu}}{x_{14}^2} - \frac{x_{13,\mu}}{x_{13}^2} \right) + v(d-2)\left(\frac{1}{x_{14}^2} - \frac{1}{x_{13}^2} \right) \right] \nonumber \\
&+ 2\mu \left[\partial^\mu u \left(\frac{x_{12,\mu}}{x_{12}^2} - \frac{x_{13,\mu}}{x_{13}^2} \right) + u(d-2)\left(\frac{1}{x_{12}^2} - \frac{1}{x_{13}^2} \right) \right] -(\alpha + \gamma')\left(\lambda x_{12}\cdot \partial v + \mu x_{12}\cdot \partial u \right) \nonumber \\
&+ 2\lambda v \left(\frac{x_{12}\cdot x_{13}}{x_{13}^2} + \frac{x_{12}\cdot x_{14}}{x_{14}^2} \right) + 2\mu u \left(\frac{x_{12}\cdot x_{13}}{x_{13}^2} - 1 \right) \bigg\}, \\
g_2 &= g_1(\lambda \leftrightarrow \lbar, \mu \leftrightarrow \mbar), \\
g_2' &= g_1'(\lambda \leftrightarrow \lbar, \mu \leftrightarrow \mbar),  \\
h &= c_1\chi^{10} \bigg\{ \frac{2\gamma}{x_{24}^2} \left[d - x_{12}\cdot x_{14} + 2\left(\alpha \frac{x_{12}\cdot x_{14}}{x_{12}^2} + \gamma - k + k\frac{x_{14}\cdot x_{13}}{x_{13}^2} \right) + x_{14}^2\left(\alpha + (\gamma - k)\frac{x_{12}\cdot x_{14}}{x_{14}^2} + k\frac{x_{12}\cdot x_{13}}{x_{13}^2} \right) \right] \bigg\}, \\
h' &= c_2 \chi^{01} \bigg\{2(1-\alpha - \gamma')\left(\alpha + \frac{x_{12}\cdot x_{14}}{x_{14}^2}(\gamma'+k) + k\frac{x_{12}\cdot x_{13}}{x_{13}^2} \right)
 \nonumber \\
&+ x_{12}^2\left[2\left(\alpha\frac{x_{12,\mu}}{x_{12}^2} + \frac{x_{14,\mu}}{x_{14}^2}(\gamma'-k) + k\frac{x_{13,\mu}}{x_{13}^2} \right)^2 + (d-2)\left(\frac{\alpha}{x_{12}^2} + \frac{\gamma'-k}{x_{14}^2} + \frac{k}{x_{13}^2} \right) \right] - d(\alpha + \gamma') \bigg\},\end{align} where $c_1$ and $c_2$ are arbitrary, independent coefficients, $d$ refers to CFT$_d$, $u$ and $v$ are the conformally invariant cross ratios, $\partial \equiv \partial_{x_1,\mu}$, and $\chi$ is the partial wave pre-factor. If the current is conserved, one may relate $c_1$ and $c_2$ (see methods for conserved tensors in \cite{blocks}). The quantities $\alpha$, $\gamma$, $\gamma'$, $k$, are \begin{align} \alpha &\equiv \frac{\Delta_1 + \Delta_2 + 1}{2}, \\
\gamma &\equiv \frac{\Delta_1 - \Delta_2 + 1}{2},\\
\gamma' &\equiv \frac{\Delta_1 - \Delta_2 -1}{2}, \\
k &\equiv \frac{\Delta_3 - \Delta_4}{2}.  \end{align} Finally, the functions $\mu$, $\mbar$, $\lambda$, and $\lbar$ are defined in Eqs. (\ref{eqn:mu1}) - (\ref{eqn:lbar}).

\textbf{Large $\Delta$ Limit:} \begin{align} F_1 &= \Delta^2 \left\{ \frac{1}{16 z} - \frac{1}{8z\sqrt{z}} - \frac{1}{4z^2(z-\zbar)}\left[ \frac{1}{4\sqrt{z}} \left(z+ \zbar \right) + \left( \zbar - z \right) \right] \right\} + \dots, \\
F_2 &= \Delta^2 \left(\frac{1}{4z\zbar} + \frac{1}{8 z \sqrt{\zbar}} + \frac{1}{16 z\zbar \sqrt{z}} + \frac{1}{32 z \sqrt{z\zbar}} \right) + \dots\\
F_3 &= \Delta^2\left( \frac{5}{16 z \zbar} + \frac{1}{4 z\sqrt{\zbar}} \right) + \dots,\end{align} where the $\dots$ indicate terms that are order $\Delta$ and below.

\bibliographystyle{utphys}
\bibliography{biblio}

\end{document}